\documentclass{aa}
\usepackage{txfonts}
\usepackage{natbib}
\usepackage{amssymb}
\bibpunct{(}{)}{;}{a}{}{,}
\usepackage{epsf}
\usepackage{graphicx}
\usepackage{epsfig}
\usepackage{graphics}
\usepackage{subfigure}
\usepackage{longtable}
\usepackage[english]{babel}
\usepackage{rotating}
\usepackage{color}

\newcommand{\teff}{$T_{\rm{eff}}$}

\newcommand{\lL}{\ifmmode \log \frac{L}{L_{\sun}} \else $\log \frac{L}{L_{\sun}}$\fi}
\newcommand{\mdot}{$\dot{M}$}
\newcommand{\myr}{M$_{\sun}$ yr$^{-1}$}
\newcommand{\vsini}{$V$~sin$i$}
\newcommand{\vinf}{$v_{\infty}$}

\newcommand{\vmac}{$v_{\rm mac}$}

\newcommand{\kms}{km~s$^{-1}$}
\newcommand{\msun}{M$_{\sun}$}

\begin{document}

\title{Spectroscopic evolution of massive stars on the main sequence}
\author{F. Martins\inst{1}
\and A. Palacios\inst{1}
}
\institute{LUPM, Universit\'e de Montpellier, CNRS, Place Eug\`ene Bataillon, F-34095 Montpellier, France  \\
%           \email{fabrice.martins@umontpellier.fr}
}

\offprints{Fabrice Martins\\ \email{fabrice.martins@umontpellier.fr}}

\date{Received / Accepted }

\abstract
{The evolution of massive stars depends on several parameters, and the relation between different morphological types is not fully constrained.}
{We aim to provide an observational view of evolutionary models in the Hertzsprung--Russell diagram, on the main sequence. This
view should help compare observations and model predictions.}
{We first computed evolutionary models with the code STAREVOL for initial masses between 15 and 100 \msun. We subsequently calculated atmosphere models at specific points along the evolutionary tracks, using the code CMFGEN. Synthetic spectra obtained in this way were classified as if they were observational data: we assigned them a spectral type and a luminosity class. We tested our spectral classification by comparison to observed spectra of various stars with different spectral types. We also compared our results with empirical data of a large number of OB stars.}
{We obtain spectroscopic sequences along evolutionary tracks. In our computations, the earliest O stars (O2-3.5) appear only above $\sim$50 \msun. For later spectral types, a similar mass limit exists, but is lower. A luminosity class V does not correspond to the entire main sequence. This only holds for the 15 \msun\ track. As mass increases, a larger portion of the main sequence is spent in luminosity class III. Above 50 \msun, supergiants appear before the end of core-hydrogen burning. Dwarf stars (luminosity class V) do not occur on the zero-age main sequence above 80 \msun. Consequently, the distribution of luminosity class V in the HR diagram is not a diagnostic of the length of the main sequence (above 15 \msun) and cannot be used to constrain the size of the convective core. The distribution of dwarfs and giants in the HR diagram that results from our calculations agrees well with the location of stars analyzed by means of quantitative spectroscopy. For supergiants, there is a slight discrepancy in the sense that luminosity class I is observed slightly earlier (i.e., at higher \teff) than our predictions. This is mainly due to wind densities that affect the luminosity class diagnostic lines. 
We predict an upper mass limit for dwarf stars ($\sim$60 \msun) that is found consistent with the rarity of O2V stars in the Galaxy. Stars with WNh spectral type are not predicted by our models. Stronger winds are required to produce the characteristic emission lines of these objects. 
}
{}


\authorrunning{Martins F. \& Palacios A.}
\titlerunning{Spectroscopic evolution of massive stars}

\maketitle

%%%%%%%%%%%%%%%%%%%%%%%%%%%%%%%%%%%%%%%%%%%%%%%%%%%%%%%%%%%%%%%%%%%%%%%%%%%%%%%%%%%%%%%%%%%%%%%%%%%%%%%%%%%%%%%%%%%%%%%%%%%%%%%
%%%%%%%%%%%%%%%%%%%%%%%%%%%%%%%%%%%%%%%%%%%%%%%%%%%%%%%%%%%%%%%%%%%%%%%%%%%%%%%%%%%%%%%%%%%%%%%%%%%%%%%%%%%%%%%%%%%%%%%%%%%%%%%
\section{Introduction}
\label{s_intro}

The evolution of massive stars depends on several parameters. The initial mass is the main parameter. Mass loss and rotation are two other important ingredients \citep{cm86,mm00}. Mass loss affects the stellar mass at all times of the evolution and shapes their path in the Hertzsprung-Russell (HR) diagram. Mass and luminosity are tightly coupled. Rotation modifies the shape and internal structure of stars and triggers mixing processes that transport chemical species and angular momentum. Evolutionary tracks are affected \citep{mema00,brott11a}. Metallicity plays an important role as well because of its effects on internal structure and mass loss: stars with a lower metallicity have reduced mass-loss rates \citep{mokiem07}. Finally, the presence of a companion can affect the evolution of massive stars by interactions \citep{langer12}. 

Stellar evolution models are used to study the effects of these various parameters on the temporal evolution of internal and external properties of massive stars. To constrain their physics, they need to be compared with results from the analysis of stars in various environments. Studies based on quantitative spectroscopy using atmosphere models are especially relevant since they provide the main physical properties (effective temperature, surface gravity, luminosity, and mass-loss rates) from observations, usually spectroscopy \citep[e.g.,][]{kud89,her92,hil03,rep04,dufton06,mokiem07,mark08,ngc2244,arias16}. This process of constraining stellar evolution models from results of quantitative analysis is necessary because the evolutionary calculations do not predict spectral characteristics. Instead, they provide the fundamental physical properties of stars. A large number of observations is thus required to draw general conclusions regarding the sequence of evolutionary phases that
are followed by stars of different initial mass, metallicity, and rotation rates. Covering the entire parameter space is difficult and currently only leads to a patchy ``observational'' picture of stellar evolution. For instance, the relations between different types of stars are described qualitatively by the Conti scenario \citep{conti75}, which relates O-type and Wolf-Rayet stars. However, the details remain largely unknown. If general trends exist, as summarized by \citet{paul07}, for instance, very few detailed empirical sequences of evolution exist. \citet{massey00,massey01} performed an observational study of clusters and OB associations in the Magellanic Clouds and the Galaxy. They provided evidence that a metallicity-dependent mass threshold exists for OB stars to enter the Wolf-Rayet phase. \citet{paul97} were the first to establish an evolutionary sequence between specific spectral types: they showed that O8If supergiants evolve into WN9ha stars. \citet{gc07} and \citet{arches} defined sequences between WN8--WN8/WC9--WC9 and O4-6I--WN8-9h stars, respectively, from observations of the Galactic center and the Arches clusters. Such spectroscopic sequences remain rare, but will most likely emerge in the near future from the studies of large samples of massive stars.

Since providing empirical spectroscopic sequences of evolution is still difficult, we can attempt to provide ``theoretical'' sequences by computing the spectral appearance of evolutionary models. A major attempt in this direction was presented in \citet{costar1,costar2} and \citet{costar3}. The authors used atmosphere models to compute synthetic spectra at specific positions along evolutionary tracks. They provided spectral energy distributions, ionizing fluxes, and ultraviolet (UV) spectra throughout the HR diagram. However, the limited quality of the atmosphere models available at that time did not allow an accurate prediction of spectral lines, especially those in absorption. In particular, it was not possible to quantitatively predict the evolution of spectral types and luminosity classes, and thus to pinpoint the relation between various classes of massive stars. A subsequent effort in this direction was made by \citet{groh14}, who adopted the same method as \citet{costar1}, but with advanced non-LTE atmosphere models including a full treatment of line-blanketing and proper radiative transfer with detailed line-broadening prescriptions. Groh et al. could thus predict the spectroscopic evolution of a 60 \msun\ star from the zero-age main sequence (ZAMS) to the supernova explosion. 

In parallel to the study of \citet{groh14}, we developed a project with the same aim of predicting the spectroscopic appearance of massive stars along their evolution for different parameters (initial masses, rotation, metallicity, and mass loss). In a first paper, we compared evolutionary tracks from different codes and groups in order to identify the uncertainties in the predictions of such models, and to choose the best models for further calculations \citep{mp13}. The general conclusion was that different models agree reasonably well on the main sequence (MS), but strongly disagree in later phases. In this paper, we thus present our calculation of the spectroscopic appearance of these stars. We focus on stellar evolution without rotation, but cover a wide mass range (15 to 100 \msun) at solar metallicity. Our aim is to provide a theoretical spectroscopic view of the evolution
of massive stars. 

The paper is organized as follows. In Sect.\ \ref{s_mod} we explain how we conducted our calculations and how we proceeded for the spectral classification of our models. Section\ \ref{s_res} presents our results. They are discussed and compared to observations and previous studies in Sect.\ \ref{s_disc}. We give our summary and conclusions in Sect.~\ref{s_conc}.

%%%%%%%%%%%%%%%%%%%%%%%%%%%%%%%%%%%%%%%%%%%%%%%%%%%%%%%%%%%%%%%%%%%%%%%%%%%%%%%%%%%%%%%%%%%%%%%%%%%%%%%%%%%%%%%%%%%%%%%%%%%%%%%
\section{Synthetic spectra and classification}
\label{s_mod}

In this section we describe how we obtained synthetic spectra from evolutionary and atmosphere models. We subsequently summarize the criteria we used for spectral classification.

%--------------------------
\subsection{Evolutionary and atmosphere models}
\label{s_evatm}

\begin{figure}[]
\centering
\includegraphics[width=0.49\textwidth]{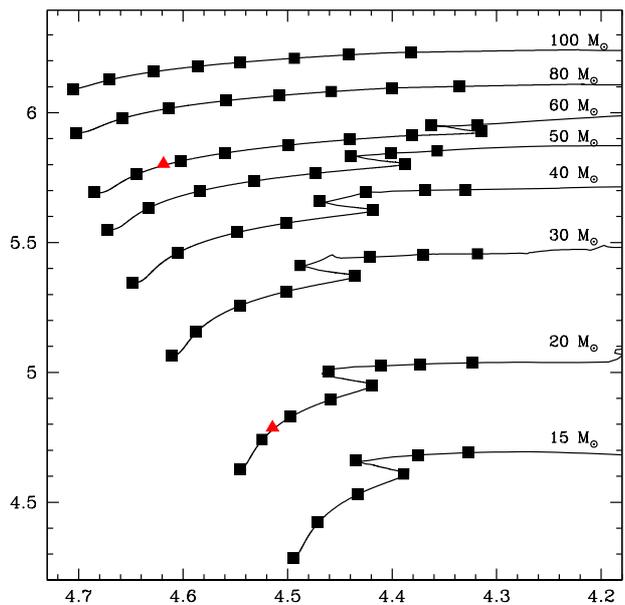}
\caption{Hertzsprung-Russell diagram showing our evolutionary tracks with solid lines and the location of the models for which a synthetic spectrum was computed (squares). The triangles highlight the position of the two models for which the temperature structure is shown in Fig.~\ref{fig_struc}.}
\label{hr_svol}
\end{figure}

We computed a set of standard non-rotating stellar evolution models with initial masses of 15, 20, 30, 40, 50, 60, 80, and 100 \msun\ at solar metallicity using the STAREVOL code \citep{dmp09,amard16}. We used the \citet{asplund09} reference values for the solar chemical mixture, and adopted Z = 0.013446 as the solar metallicity. The convective instability is treated following the mixing length formalism and the Schwarzschild criterion, with an adopted mixing-length parameter $\alpha_{\rm MLT} = 1.6304$ that has been calibrated on the Sun. A small core-overshoot ($\alpha_{\rm ov} = 0.1 H_p$,
where $H_p$ is the pressure scale height) is included in all models. 

For the treatment of mass-loss, which is crucial in determining the evolutionary path of the model stars, we used the prescription from \citet{vink01}. For models that experience a Wolf-Rayet phase (e.g., with log$_{10}$(T$_{\rm eff}$) > 4 and $X_{\rm surf} \leq 0.4$), the recipe of \citet{nl00} was adopted in these advanced phases. These phases are not discussed in the present study. In order to take clumping into account, we divided the resulting mass-loss rates by a factor of three. Several studies indicate that the true mass-loss rates of OB stars are lower than previously derived, and lower than the recipe of \citet{vink01}. \citet{mokiem07} determined mass-loss rates from H$\alpha$. They showed that a reduction of the empirical mass-loss rates by a factor two was needed in order to bring the observed modified relation of wind momentum - luminosity in agreement with the predicted relation, which makes use of the Vink et al.\ recipe. Such a reduction would be compatible with a clumping factor of about four (corresponding to a volume-filling factor of 1/4). Performing a combined H$\alpha$ and UV analysis of O supergiants, \citet{jc12} concluded that the empirical mass-loss rates were 1.5 to 2.5 times lower than the theoretical values obtained from the recipe of \citet{vink01}. \citet{surlan13} found similar reductions (factor 1.3 to 2.6) through an analysis of H$\alpha$ and UV lines (including \ion{P}{v}~1118-1128) of five O supergiants. \citet{so11} concurred to a reduction of about a factor two compared to theoretical values, although these results are mitigated by \citet{so14}. \citet{cohen14} determined mass-loss rates of OB stars from X-ray spectroscopy and obtained values reduced by a factor three (on average) compared to the predictions of \citet{vink01}. Their fitting process was able to correctly reproduce H$\alpha$ provided a clumping factor of about 20 was used.\\
In view of these results, we decided to adopt mass-loss rates reduced by a factor of three relative to the Vink et al. values in our evolutionary calculations. As such, our study is placed in the context of a maximum reduction of mass-loss rates compared to \citet{vink01}. As described below, clumping is included in the atmosphere models we compute.

In the evolutionary calculations, the atmosphere, described using the Eddington approximation, is treated as an outer boundary condition to the set of internal stellar structure equations.

\smallskip

After we calculated the evolutionary tracks, we selected points along them to compute atmosphere models. For each point, the evolutionary tracks predict the surface properties. We used the luminosity and effective radius, corresponding to an optical depth of 2/3, to calculate the effective temperature. We then computed an atmosphere model with this \teff, luminosity, and radius as input. The surface gravity, mass-loss rate, and surface chemical composition were also taken from the evolutionary model. For the wind terminal velocity (\vinf), we used \vinf\ $= 3.0 \times\ v_{\rm esc}$ where $v_{\rm esc}$ is the escape velocity \citep[e.g.,][]{garcia14}. The parameters of the atmosphere models we computed are given in Table \ref{tab_mod}. The points to which they correspond on evolutionary tracks are shown in Fig.~\ref{hr_svol}.

\begin{figure*}[t]
\centering
\includegraphics[width=0.47\textwidth]{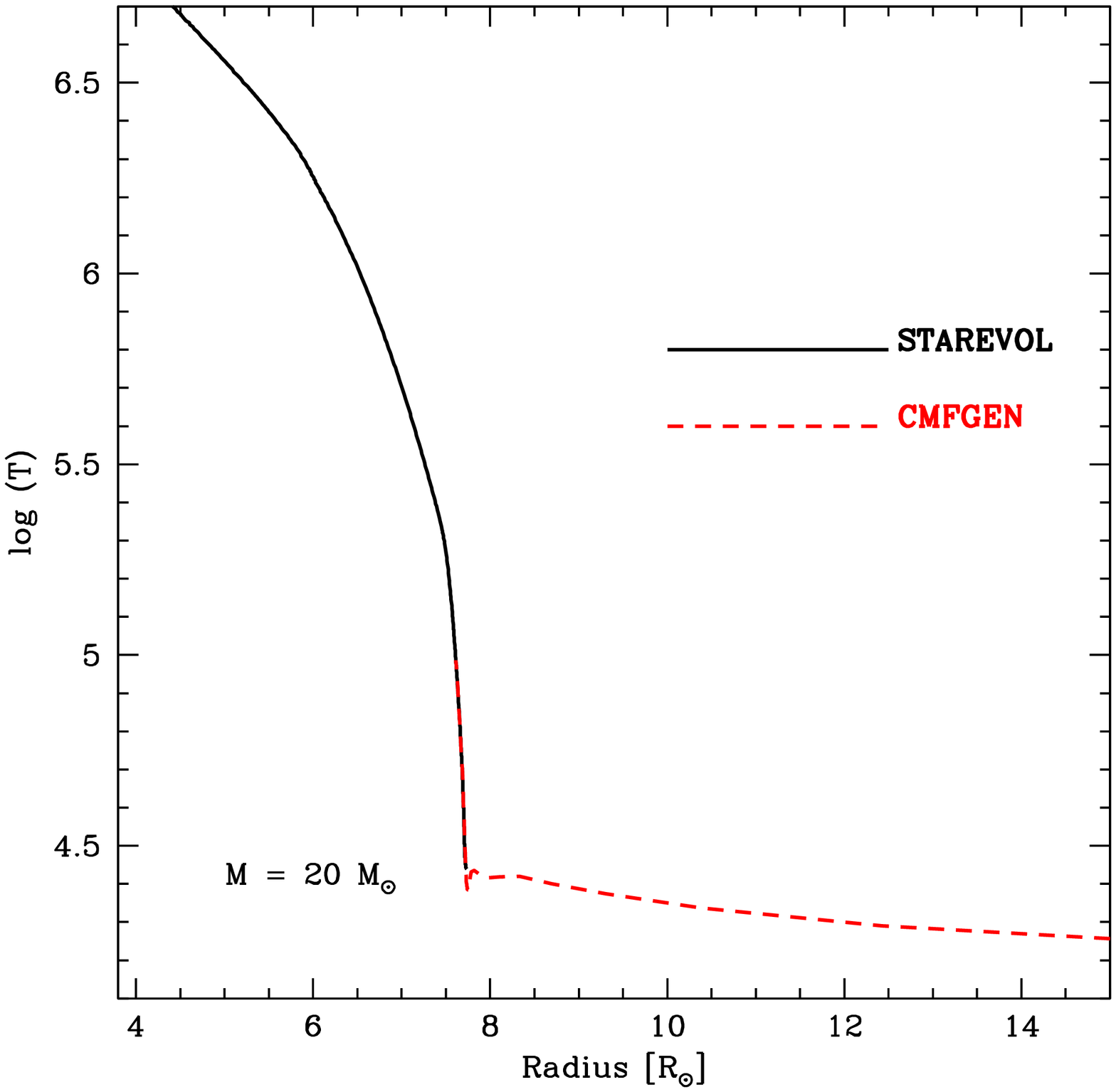}
\includegraphics[width=0.47\textwidth]{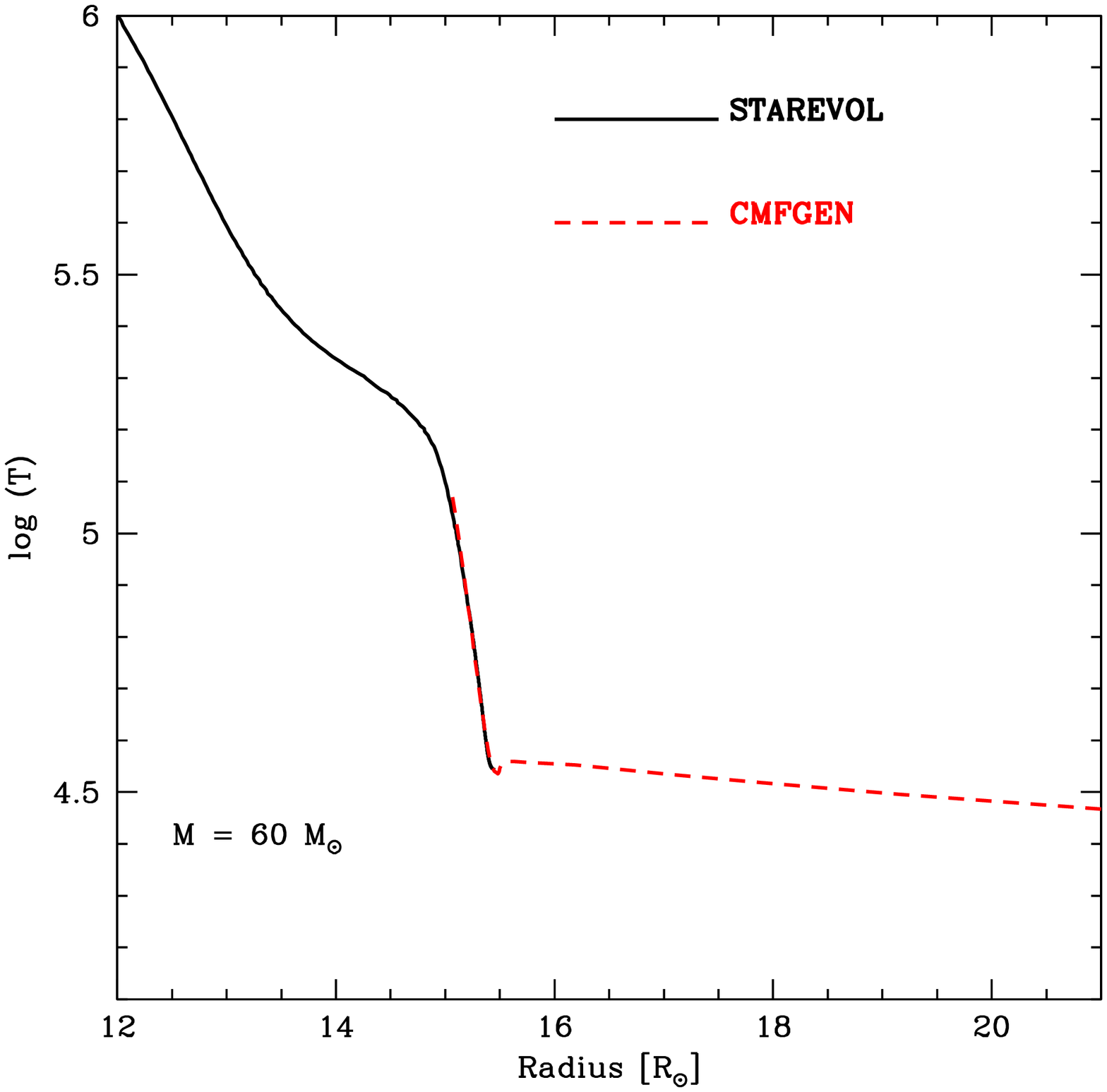}
\caption{Temperature as a function of radius in the evolutionary model (black solid line) and the atmosphere model (red dashed line) for representative points of the main sequence. The left (right) panel is for a model with $M=20 (60)$ \msun\ .}
\label{fig_struc}
\end{figure*}

The atmosphere models were computed with the code CMFGEN \citep{hm98}. CMFGEN solves the radiative transfer and rate equations iteratively, leading to non-LTE atmosphere models. A spherical geometry is adopted, and winds are included. A pseudo-photospheric velocity structure, resulting from a solution of the dynamical equations including the contribution of radiative force, is connected to a $\beta$ velocity law of the form $v = v_{\infty} (1-\frac{r}{R})^{\beta}$ - where $R$ is the photospheric radius. We adopted $\beta$ = 1.0 in our calculations. We included the following elements in the atmosphere models: H, He, C, N, O, Ne, Mg, Si, S, Ar, Ca, Fe, and Ni. 
Clumping was included by means of a volume-filling factor ($f$) approach. In practice, it is parameterized as a function of velocity $v$ (in \kms) as follows: $f = f_{\infty} + (1-f_{\infty})e^{-\frac{v}{100}}$ , where $f_{\infty}$=0.1 is the maximum filling factor reached at the top of the atmosphere (where $v$=\vinf) and $v$ is expressed in \kms. This clumping distribution may not be fully realistic since there are both theoretical \citep{ro02,so13} and observational \citep{puls06} hints that clumping may be stronger in the inner wind than in the external regions. According to the above definition, the volume-filling factor in the H$\alpha$ line formation region - usually around 30-100 \kms, see \citet{varnarval} - is on the order 0.4-0.8. This is higher -- that is, less clumped -- than other studies, which assumed a constant clumping factor throughout the atmosphere \citep[e.g.,][]{mokiem07}, a stratification that is also different from theoretical predictions and observational constraints.
Once the atmosphere models were converged, a formal solution of the radiative transfer equation was performed to produce the theoretical synthetic spectrum corresponding to the input parameters.

To ensure that continuity between the evolutionary and atmosphere models is warranted, we show in Fig.\ \ref{fig_struc} the temperature structure for two sets of models: one along the 20 \msun\ track, another along the 60 \msun\ track. They correspond to models with a central hydrogen mass fraction of about 0.4, placing them on the first half of the main sequence in the HR diagram (see Fig.\ \ref{hr_svol}). For each set, the temperature structure in the atmosphere model matches the structure in the evolutionary model
well, ensuring consistency in the calculations. Additional tests indicate that at higher luminosities and lower effective temperatures (i.e., closer to the Eddington limit) the agreement is not as good. This is a limitation of our approach (see Sect.\ \ref{slim}).

\begin{table}[ht]
\begin{center}
\caption{Atmosphere model parameters and associated spectral types.} \label{tab_mod}
\begin{tabular}{lcccccl}
\hline
M        &  T$_{\rm eff}$  &  $log(L/L_{\odot})$ &  $log g$ & log $\dot{M}$ & $v_{\infty}$ & Spectral \\    
M$_{\odot}$  &  K          &                   &          &               & \kms       & Type \\
\hline
100& 50727 &    6.09 &  4.12   & -5.62   & 3887 &   O2III/If* \\            
   & 46798 &    6.13 &  3.93   & -5.50   & 3416 &   O2III/If* \\            
   & 42444 &    6.16 &  3.72   & -5.42   & 2979 &   O3If* \\          
   & 38509 &    6.18 &  3.52   & -5.40   & 2616 &   O4If \\          
   & 35131 &    6.19 &  3.35   & -5.43   & 2370 &   O5.5-6Iaf \\       
   & 31161 &    6.21 &  3.13   & -5.43   & 2085 &   O7.5Iaf \\       
   & 27658 &    6.22 &  2.90   & -5.70   & 1803 &   O8-9Ib(f) \\       
   & 24119 &    6.23 &  2.64   & -4.70   & 1529 &   B0Ia+ \\
\hline
80 & 50390  &   5.92  &  4.18  &  -5.86  & 3858   &  O2III(f*)  \\           
   & 45514  &   5.98  &  3.93  &  -5.69  & 3261   &  O3I(f*)  \\                           
   & 41059  &   6.02  &  3.71  &  -5.61  & 2842   &  O4If         \\           
   & 36184  &   6.05  &  3.45  &  -5.60  & 2406   &  O5.5If  \\                    
   & 32201  &   6.07  &  3.23  &  -5.68  & 2112   &  O7.5Iaf  \\                   
   & 28727  &   6.08  &  3.01  &  -5.82  & 1842   &  O8-9Iab(f)  \\                   
   & 25149  &   6.10  &  2.76  &  -6.06  & 1581   &  B0-0.2Ia  \\                   
   & 21658  &   6.10  &  2.49  &  -4.91  & 1345   &  B0.7Ia+  \\                   
\hline                                      
60 & 48426  &   5.69  &  4.21  &  -6.17  & 4271   &  O3V((f*))    \\     
   & 44072  &   5.76  &  3.97  &  -6.00  & 3662   &  O3.5III((f*))     \\                
   & 40048  &   5.81  &  3.75  &  -5.90  & 3189   &  O5III(f)        \\          
   & 36332  &   5.84  &  3.54  &  -5.90  & 2786   &  O6II(f)    \\       
   & 31585  &   5.88  &  3.26  &  -5.97  & 2343   &  O7.5Ib(f)    \\      
   & 27555  &   5.90  &  3.00  &  -6.14  & 2004   &  O9.5Ib  \\                    
   & 24058  &   5.91  &  2.75  &  -5.13  & 1730   &  B0-0.5Ia+  \\                 
   & 20637  &   5.93  &  2.46  &  -5.16  & 1446   &  B0.7Ia+  \\                   
   & 23030  &   5.95  &  2.63  &  -5.05  & 1378   &  B0.5Ia+  \\                   
   & 20825  &   5.95  &  2.45  &  -5.08  & 1239   &  B0.7Ia+  \\                   
\hline                                      
50 & 47004  &   5.55  &  4.23  &  -6.38  & 4180   &  O3.5V((f))    \\      
   & 42953  &   5.63  &  3.98  &  -6.19  & 3558   &  O4V-IV((f))       \\            
   & 38362  &   5.70  &  3.71  &  -6.08  & 2995   &  O5.5III(f)   \\           
   & 34079  &   5.74  &  3.46  &  -6.08  & 2562   &  O7III-II(f)     \\     
   & 29745  &   5.77  &  3.19  &  -6.20  & 2173   &  O8.5II(f)      \\     
   & 24444  &   5.80  &  2.81  &  -6.52  & 1727   &  B0.5Ib/Iab   \\           
   & 27548  &   5.83  &  2.99  &  -6.19  & 1652   &  O9.5-9.7Ib/Iab  \\        
   & 25161  &   5.84  &  2.82  &  -6.30  & 1492   &  B0Ib         \\     
   & 22748  &   5.85  &  2.64  &  -5.18  & 1346   &  B0.7Ia+  \\                   
\hline                                      
40 & 44534  &   5.34  &  4.24  &  -6.70  & 4007   &  O4V((f))   \\    
   & 40326  &   5.46  &  3.94  &  -6.46  & 3307   &  O5V((f))     \\           
   & 35342  &   5.54  &  3.63  &  -6.36  & 2731   &  O6.5IV((f))-(f)   \\     
   & 31691  &   5.58  &  3.40  &  -6.40  & 2372   &  O8III((f))   \\       
   & 26187  &   5.62  &  3.02  &  -6.63  & 1890   &  B0Ib         \\     
   & 29484  &   5.66  &  3.19  &  -6.34  & 2046   &  O9III-II  \\     
   & 26602  &   5.69  &  2.98  &  -6.44  & 1824   &  B0Ib           \\   
   & 23369  &   5.70  &  2.75  &  -5.38  & 1598   &  B0.7Iab  \\
   & 21399  &   5.70  &  2.59  &  -5.41  & 1451   &  B1Iab  \\          
\hline                                      
30 & 40788  &   5.06  &  4.24  &  -7.17  & 3262   &  O5.5V((f))  \\                   
   & 38679  &   5.16  &  4.06  &  -6.99  & 2932   &  O6V((f))      \\         
   & 35089  &   5.26  &  3.78  &  -6.83  & 2460   &  O6.5-7V((f))     \\    
   & 31721  &   5.31  &  3.55  &  -6.83  & 2141   &  O8V-III       \\         
   & 27253  &   5.37  &  3.23  &  -6.95  & 1770   &  O9.7-B0II-Ib     \\    
   & 30763  &   5.41  &  3.23  &  -6.65  & 1942   &  O8II(f)       \\         
   & 26386  &   5.44  &  3.10  &  -6.75  & 1625   &  B0-0.5III/Ib      \\  
   & 23434  &   5.45  &  2.88  &  -5.77  & 1426   &  B0.5-0.7II/Ib/Ia+   \\
   & 20811  &   5.46  &  2.67  &  -5.83  & 1455   &  B1II/Ib/Iab   \\    
\hline                     
\end{tabular}
\end{center}
\end{table}

\setcounter{table}{0}

\begin{table}[ht]
\begin{center}
\caption{Continued} \label{tab_mod}
\begin{tabular}{lcccccl}
\hline
M        &  T$_{\rm eff}$  &  $log(L/L_{\odot})$ &  $log g$ & log $\dot{M}$ & $v_{\infty}$ & Spectral \\    
M$_{\odot}$  &  K          &                   &          &               & \kms       & Type  \\
\hline    
20 & 35084  &   4.63  &  4.24  &  -7.99  & 3443   &  O8V                \\       
   & 33458  &   4.74  &  4.04  &  -7.79  & 3051   &  O8.5V              \\            
   & 31404  &   4.83  &  3.85  &  -7.68  & 2737   &  O9.5-9.7V-IV    \\      
   & 28762  &   4.90  &  3.63  &  -7.67  & 2401   &  B0.5III              \\    
   & 26287  &   4.95  &  3.41  &  -7.73  & 2095   &  B0.5-0.7III-II          \\     
   & 28904  &   5.00  &  3.53  &  -7.42  & 2249   &  B0-0.5III-II      \\ 
   & 25776  &   5.03  &  3.31  &  -7.54  & 1978   &  B0.5-0.7III-II        \\         
   & 23612  &   5.03  &  3.15  &  -7.73  & 1801   &  B0.7-1III-II  \\
   & 21030  &   5.04  &  2.94  &  -6.51  & 1593   &  B1III-II \\
\hline
15 & 31248  &   4.28  &  4.26  &  -8.86  & 3315   &  O9.7-B0V \\
   & 29594  &   4.42  &  4.03  &  -8.65  & 2857   &  B0.5V \\
   & 27070  &   4.53  &  3.76  &  -8.57  & 2430   &  B0.5-0.7V \\
   & 24492  &   4.61  &  3.51  &  -8.62  & 2099   &  B1V \\
   & 27243  &   4.66  &  3.64  &  -8.28  & 2252   &  B0.5-0.7V-IV \\
   & 23725  &   4.68  &  3.38  &  -8.48  & 1939   &  B1V-III \\        
   & 21259  &   4.69  &  3.18  &  -7.26  & 1727   &  B2II-Ib \\   
\hline                     
\end{tabular}
\end{center}
\end{table}

%--------------------------
\subsection{Spectral classification}
\label{s_classif}

\begin{figure}[t]
\centering
\includegraphics[width=0.49\textwidth]{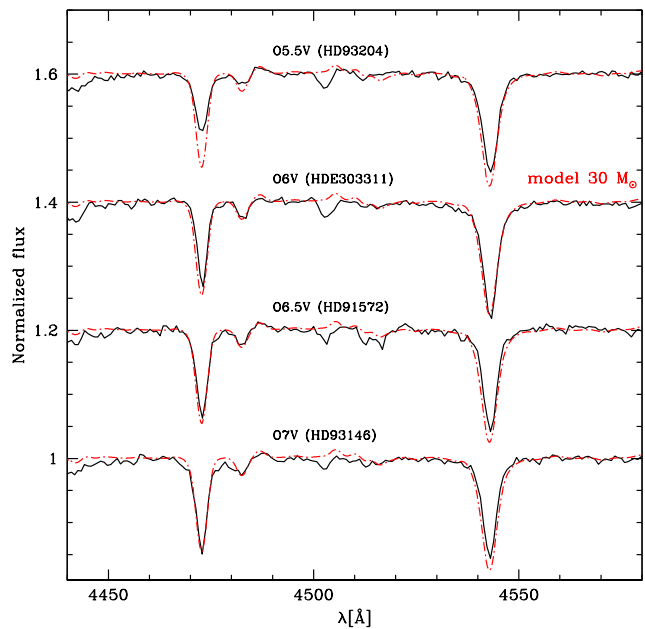}
\caption{Comparison between the $M=30$\msun\ synthetic spectrum classified as O6V((f)), see Table \ref{tab_mod}, shown by the red dot-dashed line and, from top to bottom, the GOSSS spectra of HD~93204 (O5.5V), HD~303311 (O6V), HD~91572 (O6.5V), and HD~93146 (O7V) (black solid lines). The GOSSS spectra are taken from \citet{sota11}.}
\label{sv30_goss}
\end{figure}

Before assigning a spectral type or luminosity class to the synthetic spectra, we first degraded them to an instrumental resolution of 4000. We did that to be able to compare them directly with atlases of observed spectra, which are commonly obtained at medium resolution. \\
To perform the spectral classification, we used the following standard criteria for O and B stars:

\begin{itemize}

\item \textit{spectral type}: for O4 to O9.7 stars, the ratio of the equivalent width of \ion{He}{i}~4471 to \ion{He}{ii}~4542, as defined by \citet{ca71} and \citet{mathys88}, is the prime criterion. For O8.5 to B0 stars, we added the criteria of \citet{sota11}, based on the relative strength of \ion{He}{i}~4388 to \ion{He}{ii}~4542, \ion{He}{i}~4144 to \ion{He}{ii}~4200, and \ion{Si}{iii}~4552 to \ion{He}{ii}~4542 (see their Table 4). For B stars, classification was based mainly on the relative strength of \ion{Si}{IV}~4089 to \ion{Si}{iii}~4552. Atlases of \citet{gc08}, \citet{wf90}, and \citet{evans15} were used to assess the relation between this ratio and spectral type. For the earliest O stars, we relied on the scheme defined by \citet{walborn02}, using \ion{N}{iii}~4640, \ion{N}{iv}~4058, \ion{He}{i}~4471, and \ion{He}{ii}~4542.

\item \textit{luminosity class}: the morphology of \ion{He}{ii}~4686 was used for O stars with spectral type earlier than O8.5. Detailed comparisons with the atlases of \citet{sota11,sota14} and \citet{gc08} were performed once the spectral type was assigned. For O stars with spectral types between O9 and O9.7, the relative strength of \ion{He}{ii}~4686 and \ion{He}{i}~4713 as well as the ratio \ion{Si}{iv}~4089/\ion{He}{i}~4026 were preferred, as advised by \citet{sota11}, see their Table 6. In the B-star regime, at spectral types between B0 and B0.7, luminosity classes were assigned using again the ratio \ion{Si}{iv}~4089/\ion{He}{i}~4026 together with the ratio \ion{Si}{iv}~4116/\ion{He}{i}~4121 \citep{gc08}. For later spectral types, \ion{Si}{iii}~4552/\ion{He}{i}~4388 was the main luminosity-class criterion. Following \citet{keenan71}, we assigned spectral type Ia+ to B supergiants with H$\alpha$ emission.

\end{itemize}

\begin{figure}[t]
\centering
\includegraphics[width=0.49\textwidth]{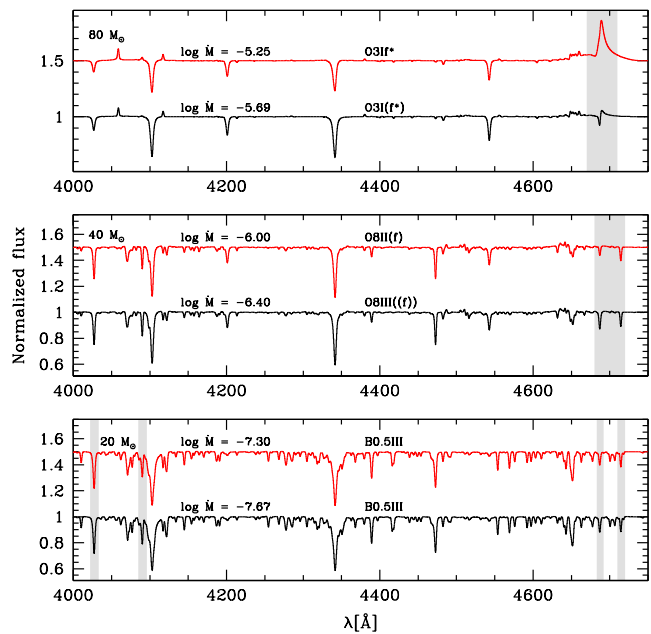}
\caption{Effect of a change in mass-loss rate on the spectral classification. In each panel, the bottom black spectrum is the initial model, while the upper red one is the same model in which the mass-loss rate has been increased by a factor $\sim$2.5 (only
in the atmosphere model). The upper (middle, lower) panel corresponds to a model with an initial mass of 80 (40, 20) \msun. More information on the models is available in Table \ref{tab_mod}. Shaded areas highlight the main luminosity-class diagnostic lines.}
\label{effect_wind_ST}
\end{figure}

Our classification scheme is mainly a \textit{\textup{qualitative}} one relying on eye inspection of line ratios and comparison to atlases. Only for the spectral types of O stars did we use the \textit{\textup{quantitative}} criterion defined by \citet{ca71} and \citet{mathys88}. We decided to proceed in this way because the quantitative scheme of Conti \& Alschuler and Mathys covers the entire range of O stars. Unfortunately, there is no such quantitative scheme for B stars. The classification criteria for luminosity classes are usually limited to narrow spectral type ranges, and quantitative criteria do not exist for all ranges. We thus decided to rely on comparisons to observed spectra in order to provide a consistent classification of luminosity classes. Using this method, we sometimes found uncertain spectral types and/or luminosity classes. In these cases we list a range of plausible values (see Table.\ \ref{tab_mod}).

To check the accuracy of our classification process, we show in Fig.\ \ref{sv30_goss} a comparison of our synthetic spectrum for the 30 \msun\ model that we classify as O6V((f)) with observed spectra taken from the GOSSS survey\footnote{Spectra from \citet{sota11}, available on this web site: \url{http://ssg.iaa.es/en/content/galactic-o-star-catalog/}}. The GOSSS spectra cover the spectral type range O5.5 to O7 and correspond to dwarf stars. Figure\ \ref{sv30_goss} clearly shows that the \ion{He}{i}~4471 line in the O5.5V spectrum is too strong compared to our synthetic spectrum. Similarly, the \ion{He}{ii}~4542 line the O7V spectrum
is too strong. The O6V and O6.5V templates better reproduce the ratio \ion{He}{i}~4471/\ion{He}{ii}~4542 that we have in our synthetic spectrum. Quantitatively, we reach the same conclusions. Calculating the value of $\log \frac{EW(4471)}{EW(4542)}$ , where EW stands for equivalent width, we find a value of -0.26 for our model and -0.32, -0.28, -0.17, and -0.14 for the O5.5V, O6V, O6.5V, and O7V template spectra, respectively. Consequently, a classification as O6V is preferred.
Our classification method is thus accurate to within one subtype. More tests of the accuracy of the classification process are shown in Sect.\ \ref{s_res}.

We also studied the effect of wind density on the spectral classification. The results of our tests are shown in Fig.\ \ref{effect_wind_ST}. We selected three models of the 20, 40, and 80 \msun\ sequences. We show the initial spectra in the bottom part of each panel. We then computed new models with the same initial parameters, except for the mass-loss rate, which was increased by a factor $\sim$2.5. The resulting spectra are shown in the upper part of each panel. Starting from the bottom panel, we see that the change in mass-loss rate for our 20 \msun\ model does not affect the classification lines. Consequently, the spectral type or luminosity class remain unchanged. For the 40 \msun\ model shown in the middle panel, the ratio \ion{He}{i}~4471/\ion{He}{ii}~4542 is not affected by variations in \mdot, but \ion{He}{ii}~4686 evolves from strong to weak absorption (the neighboring \ion{He}{i}~4713 line is useful for comparison). Hence, the spectral type remains O8, but the luminosity class shifts from III((f)) to II(f), according to the criterion of \citet{sota11}, see their Table 5. Finally, the upper panel shows the 80 \msun\ model. The relative strength of the \ion{N}{iv}~4058 and \ion{N}{iii}~4640 lines and the intensity of \ion{He}{i}~4471 are barely affected by a mass-loss rate increase. However, \ion{He}{ii}~4686 strengthens from a weak P-Cygni profile to a full emission line. Initially classified as O3I(f*) because of the ``intermediate'' nature of \ion{He}{ii}~4686, an increase in \mdot\ unambiguously translates into the classification O3If*. 

We also stress that classification of O-type stars depends on spectral resolution and rotational velocity, as shown by \citet{markova11}. Spectral types are usually more affected than luminosity classes. Differences of up to one subclass can be encountered depending on resolving power and \vsini.

%%%%%%%%%%%%%%%%%%%%%%%%%%%%%%%%%%%%%%%%%%%%%%%%%%%%%%%%%%%%%%%%%%%%%%%%%%%%%%%%%%%%%%%%%%%%%%%%%%%%%%%%%%%%%%%%%%%%%%%%%%%%%%%
%%%%%%%%%%%%%%%%%%%%%%%%%%%%%%%%%%%%%%%%%%%%%%%%%%%%%%%%%%%%%%%%%%%%%%%%%%%%%%%%%%%%%%%%%%%%%%%%%%%%%%%%%%%%%%%%%%%%%%%%%%%%%%%
\section{Results}
\label{s_res}

\begin{figure*}[]
\centering
\includegraphics[width=0.95\textwidth]{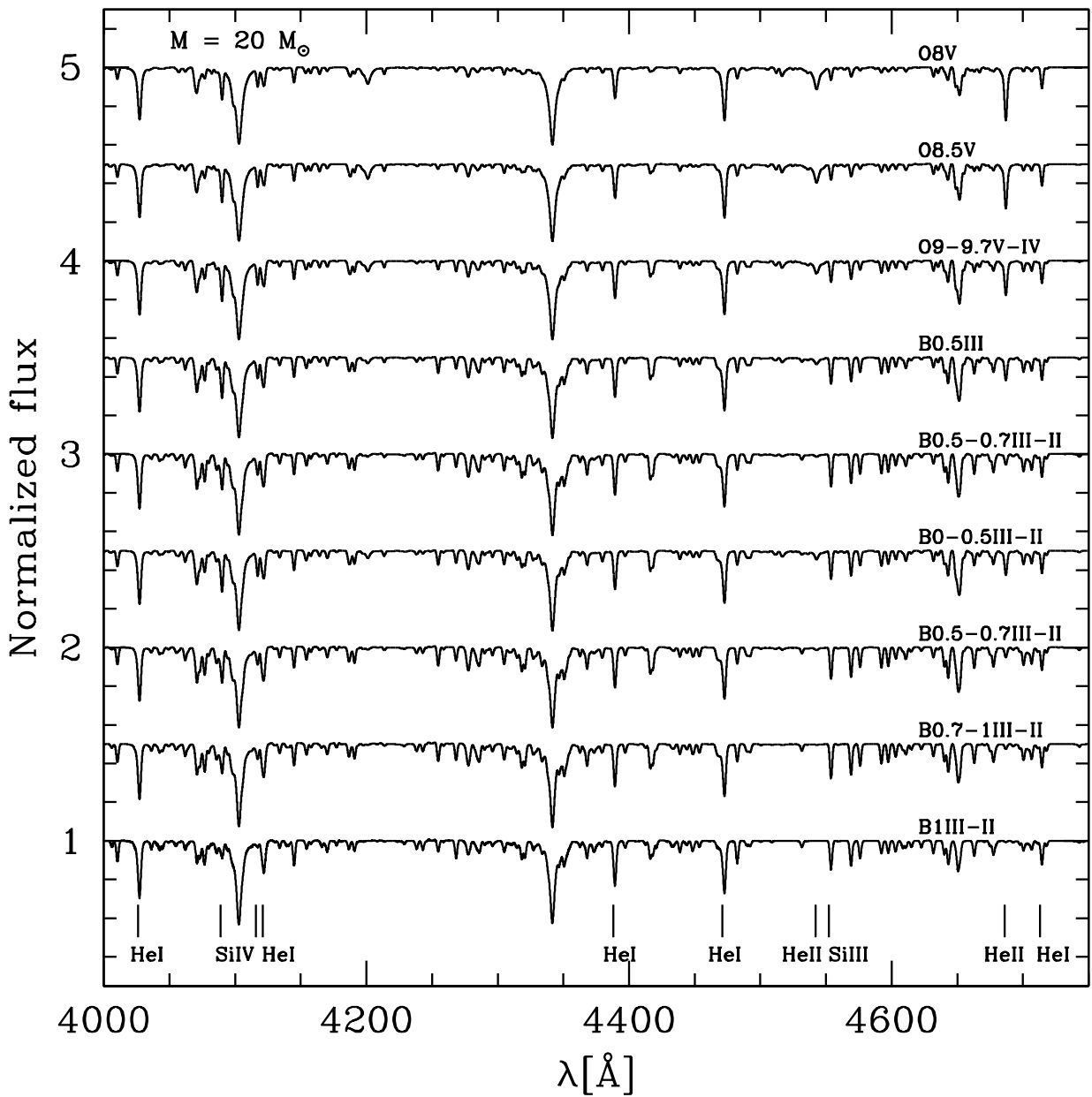}
\caption{Spectroscopic sequence along the M=20\msun\ evolutionary track, from top to bottom, between 4000 and 4750 \AA. The main classification lines are indicated. The spectra have been shifted by 0.5 for clarity.}
\label{sv20_opt}
\end{figure*}

\begin{figure*}[t]
\centering
\includegraphics[width=0.95\textwidth]{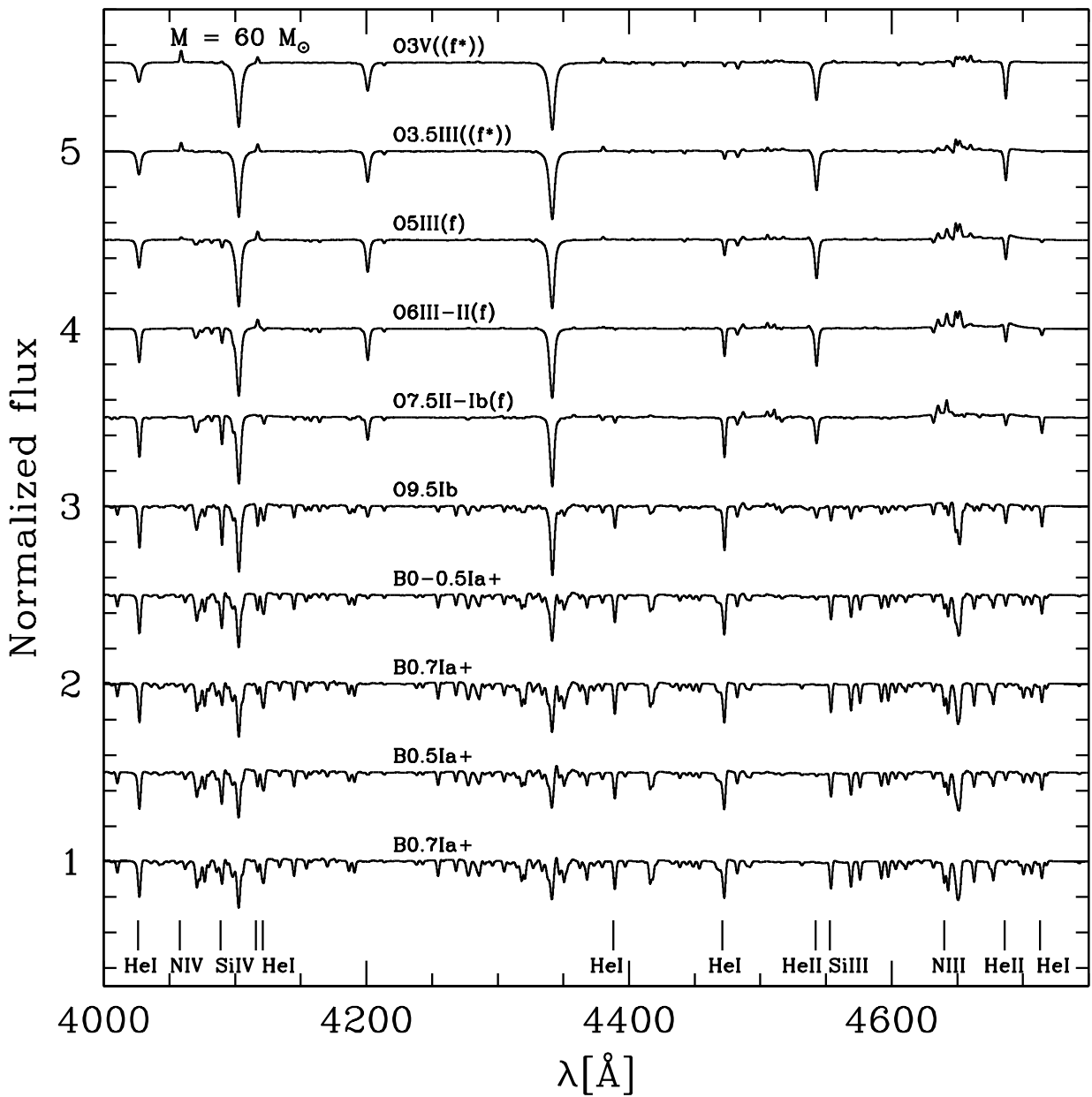}
\caption{Same as Fig.\ \ref{sv20_opt} for the M=60\msun\ models.}
\label{sv60_opt}
\end{figure*}

The results of our spectral classification are gathered in the right column of Table~\ref{tab_mod}. For each initial mass, they provide a spectroscopic evolutionary sequence. 
We show the synthetic spectra for the models along the 20 and 60 \msun\ evolutionary tracks in Fig.\ \ref{sv20_opt} and \ref{sv60_opt}, respectively. We have selected the optical range between 4000 and 4750 \AA\ since it is the most commonly used for spectral classification.

\begin{figure*}[t]
\centering
\includegraphics[width=0.47\textwidth]{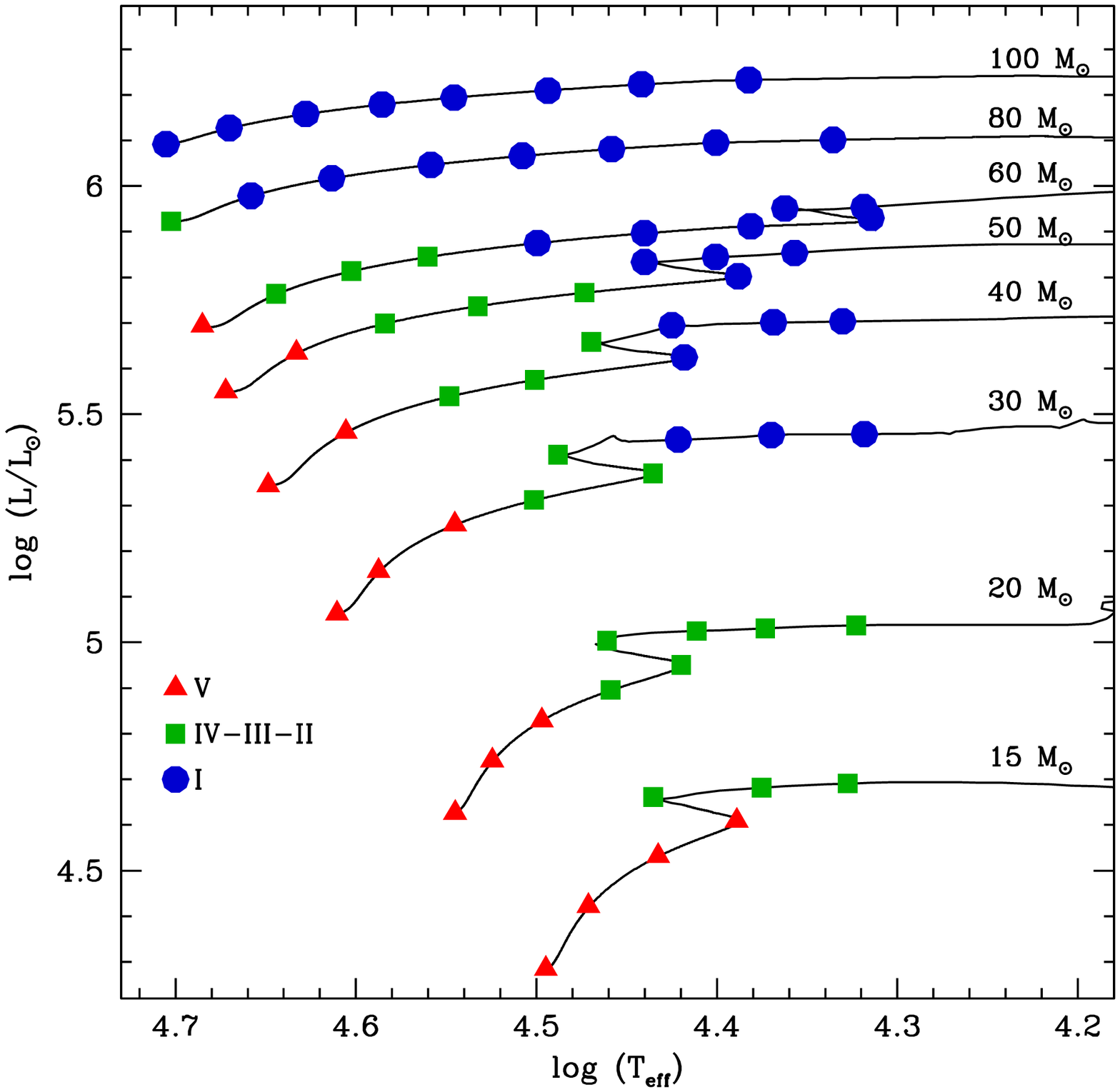}
\includegraphics[width=0.47\textwidth]{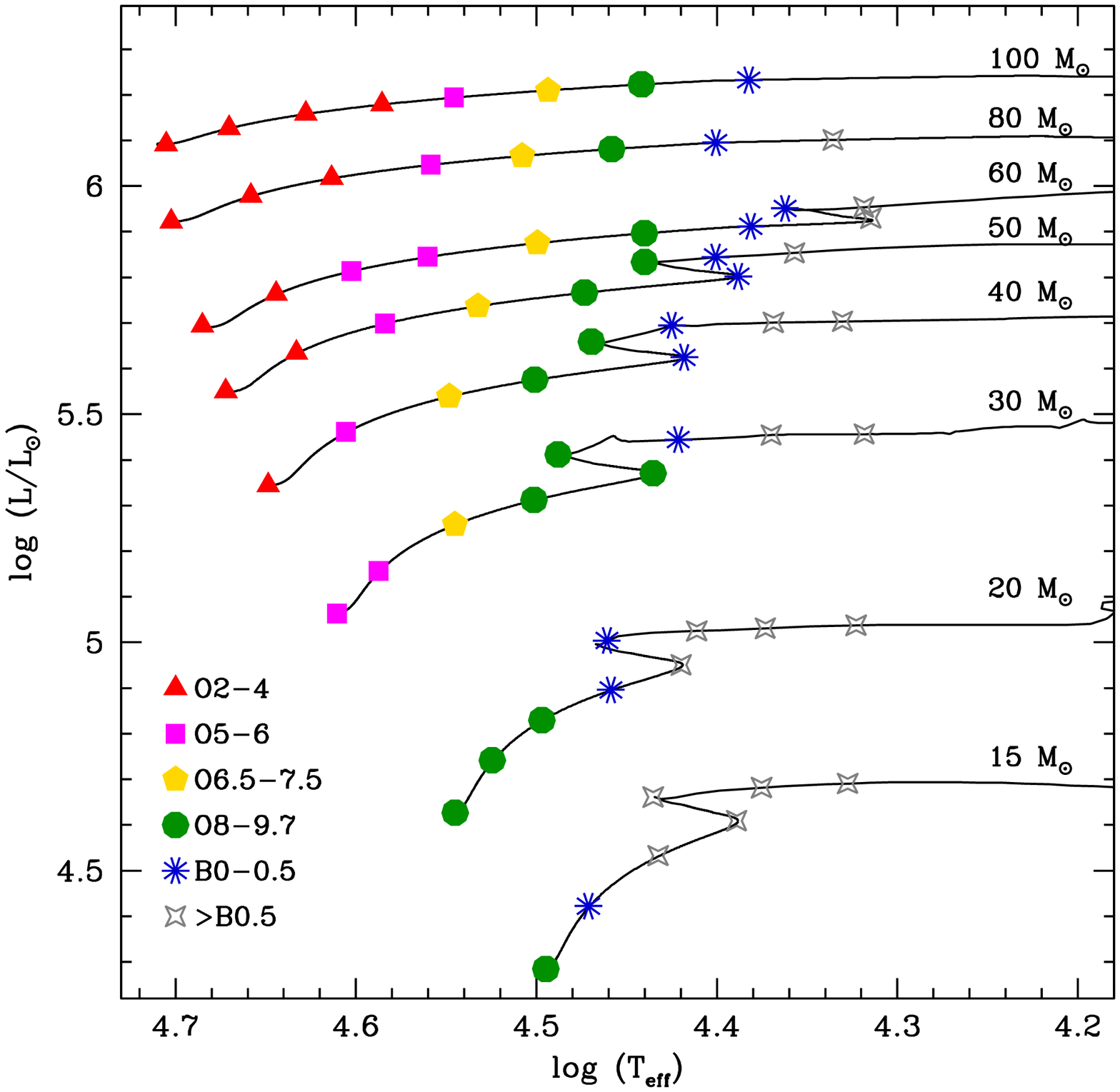}
\caption{Same as Fig.\ \ref{hr_svol}, but with different symbols corresponding to different luminosity classes (right panel) and spectral types (left panel), as inferred from the synthetic spectra.}
\label{hr_ST}
\end{figure*}

\begin{figure*}[]
\centering
\includegraphics[width=0.47\textwidth]{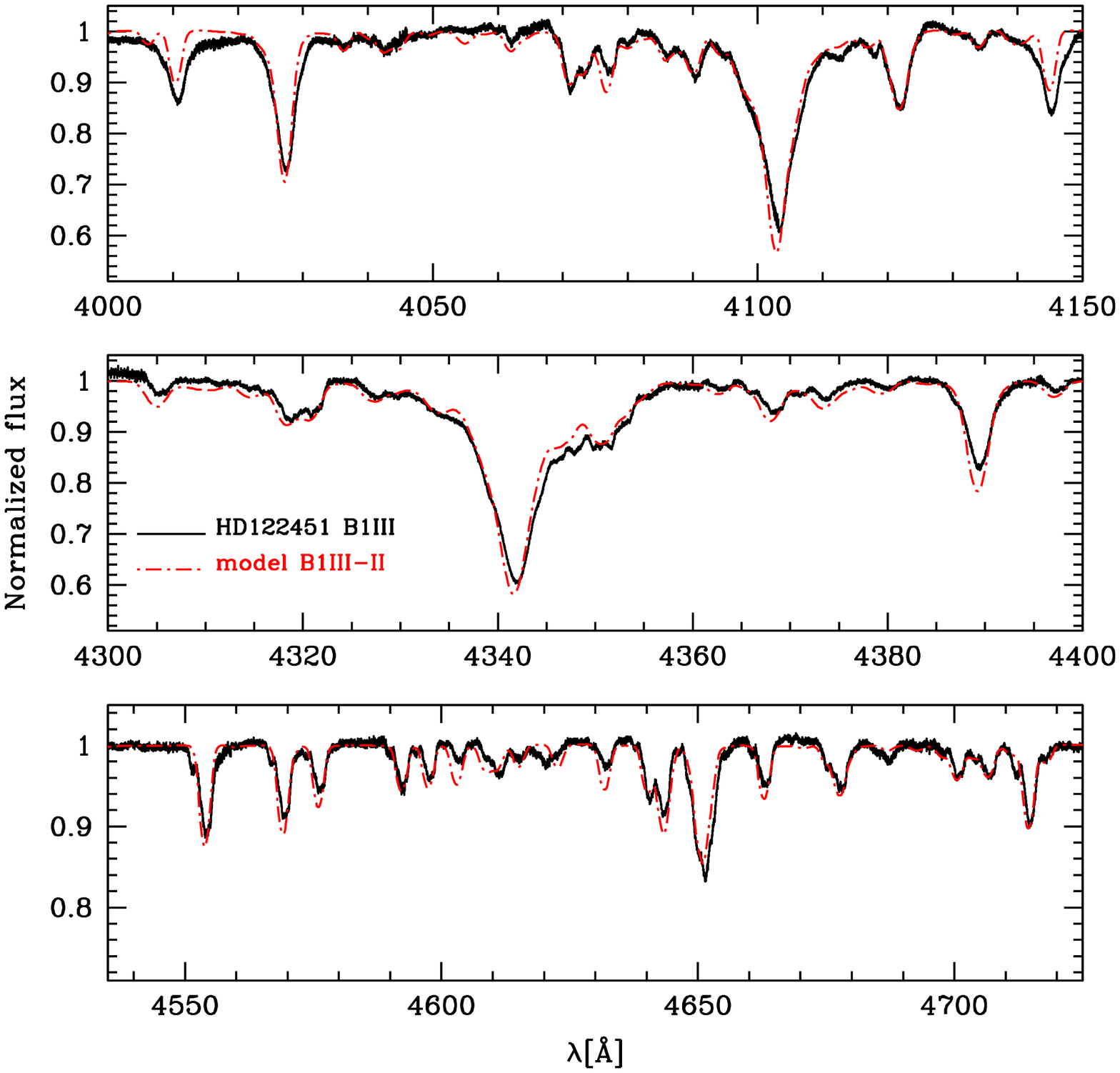}
\includegraphics[width=0.47\textwidth]{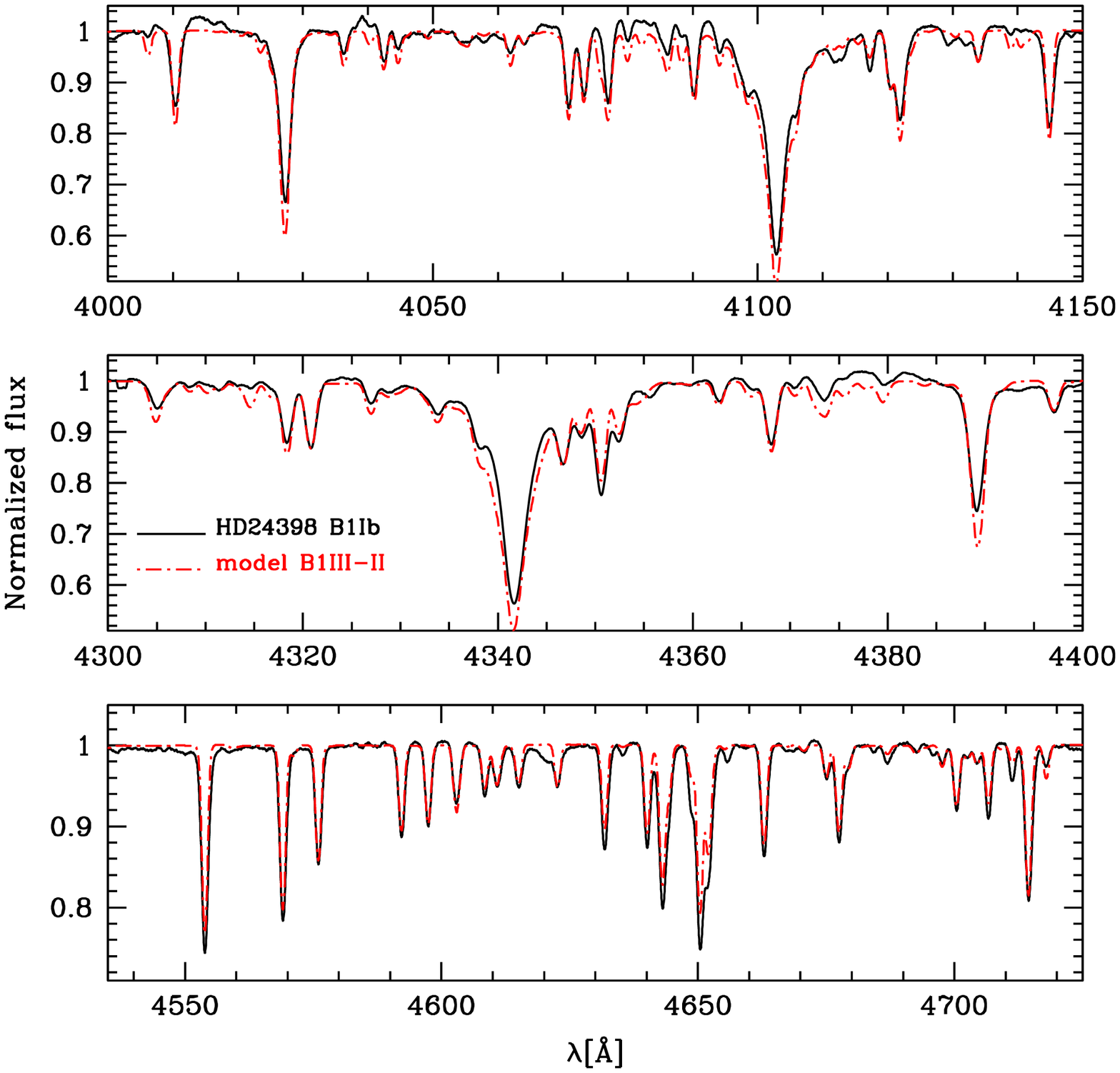}
\caption{Comparison between the synthetic spectrum of the M = 20 \msun\ star classified as B1III-II (red dot-dashed line) and the observed spectrum of HD~122451, a B1III giant (left panel),
and HD~24398, a B1Ib supergiant (right panel). The synthetic spectrum has been convolved with \vsini-\vmac\ equal to 0-150 \kms\ (left) and 50/50 \kms\ (right). }
\label{m20_I_III}
\end{figure*}

In Fig.\ \ref{sv20_opt} we see the change in ionization when moving from top to bottom, that is, away from the ZAMS: \ion{He}{i}~4471 remains roughly constant, but \ion{He}{ii}~4542 slowly disappears. This reflects the decrease in effective temperature along the evolutionary tracks. The 20 \msun\ model starts on the ZAMS as an O8 dwarf and becomes an early B giant. The complete sequence is shown in Fig.\ \ref{sv20_opt} and Table\ \ref{tab_mod} (last column). Figure\ \ref{hr_ST} displays the HR diagram shown in Fig.\ \ref{hr_svol} in which the different models have been color- and symbol-coded according to the spectral type and luminosity class of the corresponding synthetic spectrum. For an initial mass of 20 \msun, the star first appears as a dwarf, then becomes a giant or bright giant in the latest part of the main sequence, until the terminal age main sequence (TAMS) is reached. It remains a giant in the early stages of post-main-sequence evolution. To test this conclusion, we compare in Fig.\ \ref{m20_I_III} the spectrum of the 20 \msun\ star classified as B1III-II and the observed spectra of HD~122541 (a B1III giant) and HD~24398 (a B1Ib supergiant) \footnote{The spectrum of HD~122451 was shared by E. Alecian and was obtained with HARPS \citep{alecian11}. The spectrum of HD~24398 was retrieved from the ELODIE archive \citep{moul04}.}. For the purpose of the comparison, we determined the values of the projected rotational velocity (\vsini) and of the macroturbulent velocity (\vmac) using the Fourier transform of \ion{He}{i}~4713 \citep{sergio14} and the fit of that same line, respectively. The final synthetic spectrum was convolved with a rotational profile and a radial-tangential profile using these values of \vsini\ and \vmac. We stress that the goal is not to fit the observed spectrum. Instead, we wish to highlight that the ratio \ion{Si}{IV}~4089/\ion{Si}{iii}~4552 is comparable in the synthetic spectrum and both observations, confirming that they have the same spectral type (B1). The inspection of \ion{He}{i}~4388/\ion{Si}{iii}~4552 indicates that the luminosity class of our synthetic spectrum is more consistent with a giant than a supergiant. Finally, H$\gamma$ is narrower in the Ib supergiant than in our synthetic spectrum, whereas the B1III giant has a very similar line profile. We can thus safely exclude that the synthetic spectrum corresponds to a supergiant. 
In conclusion, stellar evolution along the 20 \msun\ track proceeds through luminosity classes V to III/II along the main sequence, but avoids the supergiant phase, which is not reached in the early phases after the TAMS. We also note that the main sequence does not entirely correspond to dwarf stars: before the TAMS, the star appears as a giant.

\begin{figure*}[t]
\centering
\includegraphics[width=0.32\textwidth]{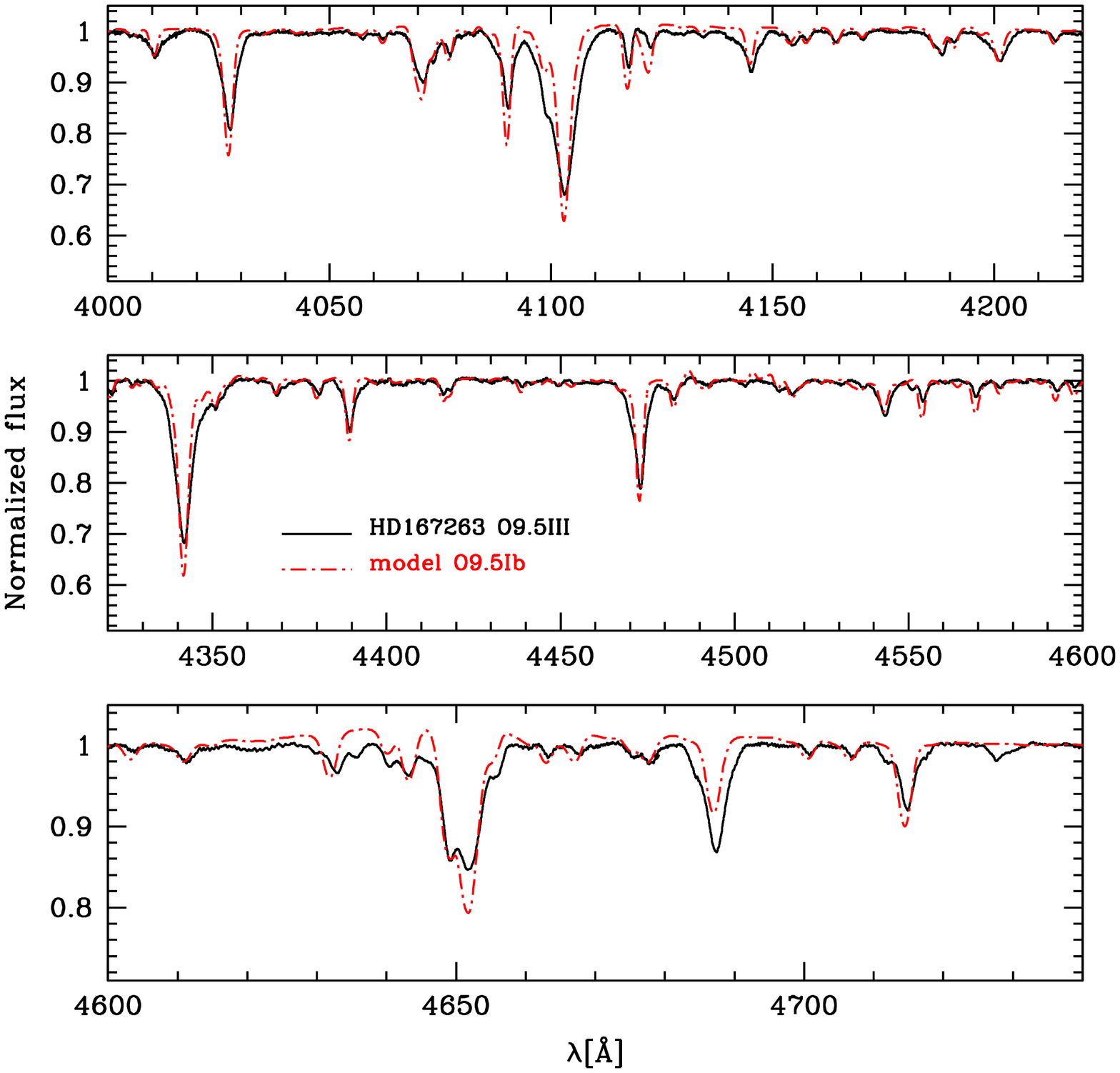}
\includegraphics[width=0.32\textwidth]{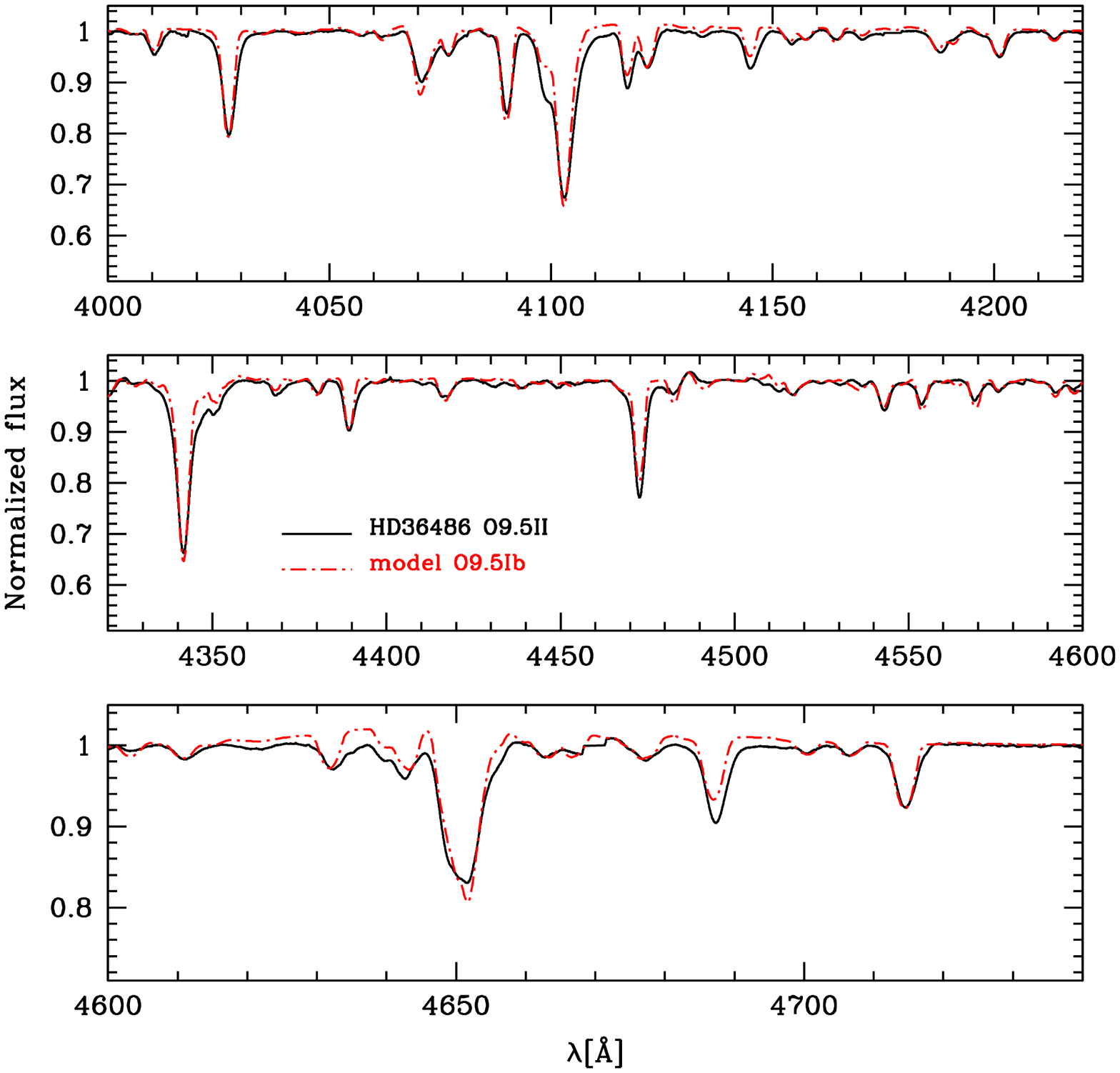}
\includegraphics[width=0.32\textwidth]{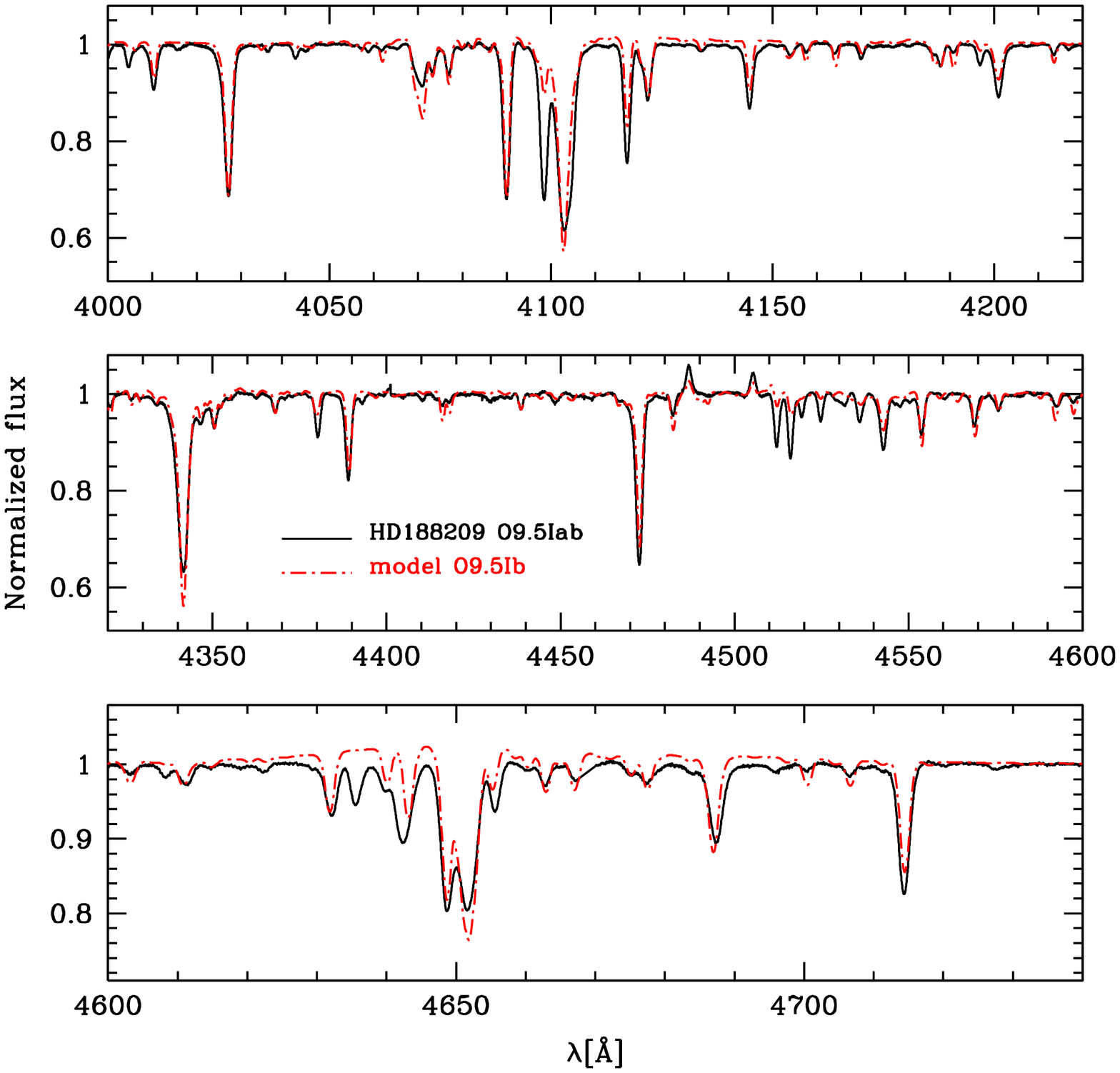}
\caption{Comparison between the synthetic spectrum of the M = 60 \msun\ star classified as O9.5Ib (red dot-dashed line) and the observed spectrum of HD~167263 a O9.5III giant (let panel), HD~36486 a O9.5II bright giant (middle panel), and HD~188209 a O9.5Iab supergiant (right panel). The synthetic spectrum has been convolved with \vsini-\vmac\ equal to 80-100 \kms\ (left), 130/60 \kms\ (middle) and 55/70 \kms\ (right).}
\label{m60_I}
\end{figure*}

Figure\ \ref{sv60_opt} shows the spectral evolution along the 60 \msun\ track. The strengthening (weakening) of \ion{He}{i}~4471 (\ion{He}{ii}~4542) with increasing distance from the ZAMS is very well seen. The 60 \msun\ model appears as an early O dwarf on the ZAMS, O3V((f*)) , but contrary to the 20 \msun\ model, it rapidly becomes a giant (see Fig.\ \ref{hr_ST}). The supergiant state is reached before the end of the main sequence, at a late-O spectral type. To test the reliability of this prediction, we show in Fig.\ \ref{m60_I} the comparison of the second 60 \msun\ spectrum appearing as a supergiant (O9.5Ib in Table \ref{tab_mod}) with observed spectra of similar spectral types. The observed spectra are the same as those used by \citet{mimesO}. In the left panel, the O9.5III giant HD~167263 is displayed using a black solid line. We first see that the \ion{He}{i}~4471 and \ion{He}{ii}~4542 lines have the same ratio in HD~167263 and the synthetic spectrum, in agreement with the spectral types: both are O9.5. However, in the observed giant star, the ratio \ion{He}{ii}~4686/\ion{He}{i}~4713 is higher than in the synthetic spectrum. According to the classification criterion defined by \citet{sota11}, this is indicative of a more evolved luminosity class than that of HD~167263, that is, of class II or I. 
In the middle panel, the bright giant HD~36486 (O9.5II) is shown. The ratio \ion{He}{ii}~4686/\ion{He}{i}~4713 is still higher in the observed spectrum than in the model, indicating that a luminosity class II is not appropriate for the synthetic spectrum. However, the difference is not as large as it is with the giant spectrum. 
Finally, in the right panel, the comparison star is HD~188209, a O9.5Iab supergiant. The \ion{He}{i}~4471/\ion{He}{ii}~4542 ratio is the same in both spectra. This time, \ion{He}{ii}~4686/\ion{He}{i}~4713 is also very similar, indicating that the luminosity class of both spectra is almost the same. We assigned a luminosity class Ib to the synthetic spectrum, while that of HD~188209 is Iab. The small difference we observe in \ion{He}{ii}~4686/\ion{He}{i}~4713 (the ratio being lower in HD~188209 than in our synthetic spectrum) is consistent with the little difference in luminosity class:  Iab versus Ib. From these comparisons, we conclude that our classification procedure is correct and that the 60 \msun\ star truly becomes a supergiant before the end of the main sequence.

Inspection of the 20 and 60 \msun\ tracks and associated spectroscopic sequences shows that luminosity class I can be reached already on the main sequence at high masses. Figure\ \ref{hr_ST} gives more information (left panel). Below 30 \msun, supergiants are not observed on the main sequence, nor immediately after. For stars with masses between 30 and 50 \msun\, , supergiants appear at the TAMS. Above 50 \msun, supergiants are encountered already on the main sequence, at locations progressively closer to the ZAMS as mass increases. Above 100 \msun, all models are classified as supergiants. 
We thus conclude that stars classified as blue supergiants are not necessarily stars that have ended core-hydrogen burning.

In the 20 \msun\ track, we have noted that close to the TAMS, stars already appear as giants. In other words, a dwarf luminosity class does not correspond to the entire main sequence. Figure\ \ref{hr_ST} indicates that below 30 \msun\ a large portion of the main sequence is spent in luminosity class V. Above this limit, dwarfs are seen at the beginning of the MS, but for a duration that is increasingly shorter as mass increases. In the 60 \msun\ model, only the first point, on the ZAMS, is classified as a dwarf. For the 80 \msun\ model, the star is already a giant on the ZAMS (see Table \ref{tab_mod}). Hence, it is clear from our investigation that the extent of the main sequence cannot be compared to the location of dwarf stars in the HR diagram. In addition, our predictions indicate that above ~80 \msun\ stars do not appear as dwarfs even in the earliest phases of their evolution (see Sect.~\ref{s_Vmax}).

The right panel of Fig.\ \ref{hr_ST} shows the location of different spectral types in the HR diagram, according to our calculations. The shift from early-O stars to early-B stars as \teff\ decreases is clearly seen. Among a given spectral type, \teff\ is lower when the luminosity is higher. This is consistent with supergiants being cooler than dwarfs \citep{vacca96,msh05}. Our results indicate that the earliest O stars (O2-O3.5) are not observed below an initial mass of about 50 \msun. For later spectral types, this minimum mass decreases quickly, such that for late-O stars (O8-9.7) a wide range of masses higher than 15 \msun\ is possible. Obviously, dwarfs have lower masses than supergiants at a given spectral type.

\begin{figure}[t]
\centering
\includegraphics[width=0.47\textwidth]{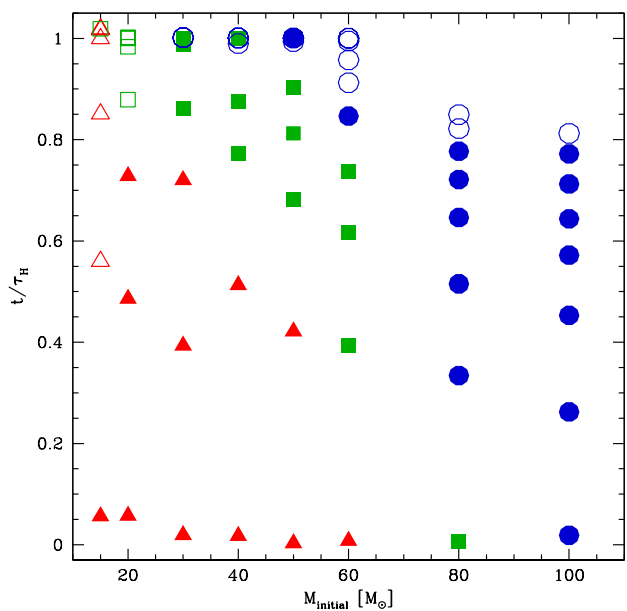}
\caption{Fraction of the time spent on the main sequence as a function of initial mass for our models. Triangles, squares, and circles correspond to dwarfs, giants, and supergiants. Filled (open) symbols indicate O (B) stars.}
\label{t_mass}
\end{figure}

Figure\ \ref{t_mass} presents our results in a different way. The fraction of the main sequence spent in different luminosity classes is shown as a function of initial mass \citep[see also Fig.~10 of ][]{langer12}. During most of the main sequence, stars with initial masses in the range 30-50 \msun\ appear as O stars. The latest part of the main sequence of stars more massive than 60 \msun\ (and of the 20 \msun\ model) is spent in the B spectral type. The 15 \msun\ model evolves most of the time as a B star. Figure\ \ref{t_mass} clearly illustrates the point we raised previously: the main sequence does not necessarily correspond to a luminosity class V. In fact, only about half of the area of the diagram that shows the 'initial mass - fraction of time spent on the main sequence' is covered by dwarfs. The dwarf phase amounts to about 80\% at 20-30 \msun. As mass increases, however, the fraction of the main sequence spent in the giant and supergiant phases increases. The 100 \msun\ model always appears as a supergiant. 

%%%%%%%%%%%%%%%%%%%%%%%%%%%%%%%%%%%%%%%%%%%%%%%%%%%%%%%%%%%%%%%%%%%%%%%%%%%%%%%%%%%%%%%%%%%%%%%%%%%%%%%%%%%%%%%%%%%%%%%%%%%%%%%
%%%%%%%%%%%%%%%%%%%%%%%%%%%%%%%%%%%%%%%%%%%%%%%%%%%%%%%%%%%%%%%%%%%%%%%%%%%%%%%%%%%%%%%%%%%%%%%%%%%%%%%%%%%%%%%%%%%%%%%%%%%%%%%
\section{Discussion}
\label{s_disc}

In this section we compare our results with the study of \citet{groh14}. We also compare our results with quantitative spectroscopic analyses of observed spectra.

%--------------------------
\subsection{Comparison with \citet{groh14}}
\label{s_compGroh}

\citet{groh14} investigated the spectral evolution along a 60 \msun\ evolutionary track computed with the Geneva code. Although we do not extend our calculations as far as they do (we stop at \teff\ $\sim$ 20000 K, i.e., just after the TAMS), we can still compare our results on the main sequence. Groh et al.\ find that their 60 \msun\ model spends 90\% of the main sequence with an O-type supergiant spectral appearance. The remaining 10\% are spent as a luminous blue variables (see their Table 1).

The first similarity with our study is that on the ZAMS, the star has a spectral type around O3. The second similarity is that the supergiant phase appears already on the main sequence. The main difference between the two studies is that the supergiant phase does not appear at the same position along the track. The reason is the different mass-loss rates. In the calculations of \citet{groh14}, the 60 \msun\ star appears as a supergiant already on the ZAMS. In our case, a luminosity class I is reached
only in the second part
of the MS. Inspection of Table 1 of \citet{groh14} shows that their first model, on the ZAMS, has \mdot\ $= 1.73 \times\ 10^{-6}$ \myr. Our first 60 \msun\ model has \mdot\ $= 6.76 \times\ 10^{-7}$ \myr, that is, lower by a factor 2.5. Taking into account the difference in clumping factor (we use 0.1, while Groh et al.\ chose 0.2 in the O-type phase) and the small difference in terminal velocity (4271 versus 3543 \kms), the difference in wind density between the two studies is of a factor $\sim$2.2 (density being lower in our study). This naturally explains the change in luminosity classes that is due to the wind density dependence of \ion{He}{ii}~4686 line (see Fig.\ \ref{effect_wind_ST}).

\begin{figure}[t]
\centering
\includegraphics[width=0.47\textwidth]{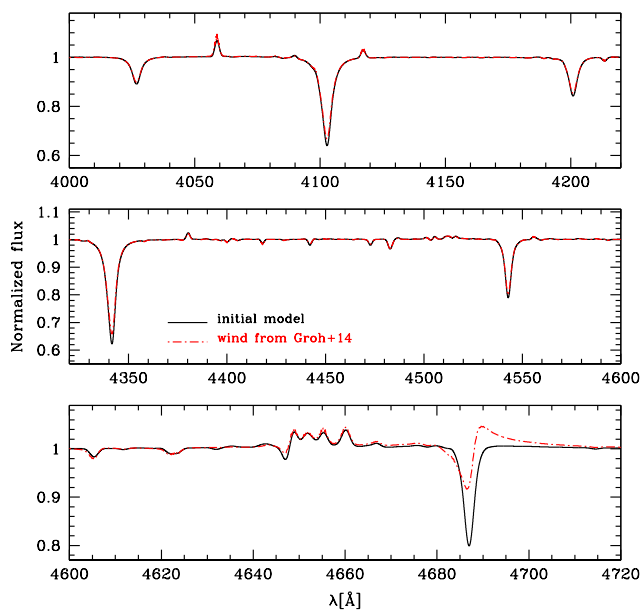}
\caption{Comparison between the synthetic spectrum of the M = 60 \msun\ ZAMS model (black solid line) and the same model with the mass-loss rate, terminal velocity, and clumping parameter of the ZAMS model of \citet{groh14}. }
\label{m60_comp_groh}
\end{figure}

To further test the difference between our calculations and those of \citet{groh14}, we show in Fig.\ \ref{m60_comp_groh} our ZAMS 60~\msun\ model together with a model computed from the same stellar parameters, but for which we used the wind parameters of the ZAMS model of Groh et al.: \mdot\ = $10^{-5.76}$ \myr, \vinf\ = 3543 \kms\ , and $f_{\infty}$ = 0.2. Most lines are basically unaffected by a change in wind density, except for the core of the Balmer lines and, to a larger extent, \ion{He}{ii}~4686. Hence, we would assign the same spectral type to the new model as in the initial model (O3). Given the P-Cygni morphology of \ion{He}{ii}~4686 in the new model, the luminosity class would be changed from V to III-I \citep[see Figs. 3 and 4 of][]{walborn02}. Inspection of Fig.~5 of \citet{groh14} reveals that their ZAMS model shows the same morphology for \ion{He}{ii}~4686. We thus conclude that the differences between our classification and theirs is rooted in the different assumptions made regarding wind parameters. 

%--------------------------
\subsection{Comparison with an observed HR diagram}
\label{s_obs_hrd}

\begin{figure*}[]
\centering
\includegraphics[width=0.49\textwidth]{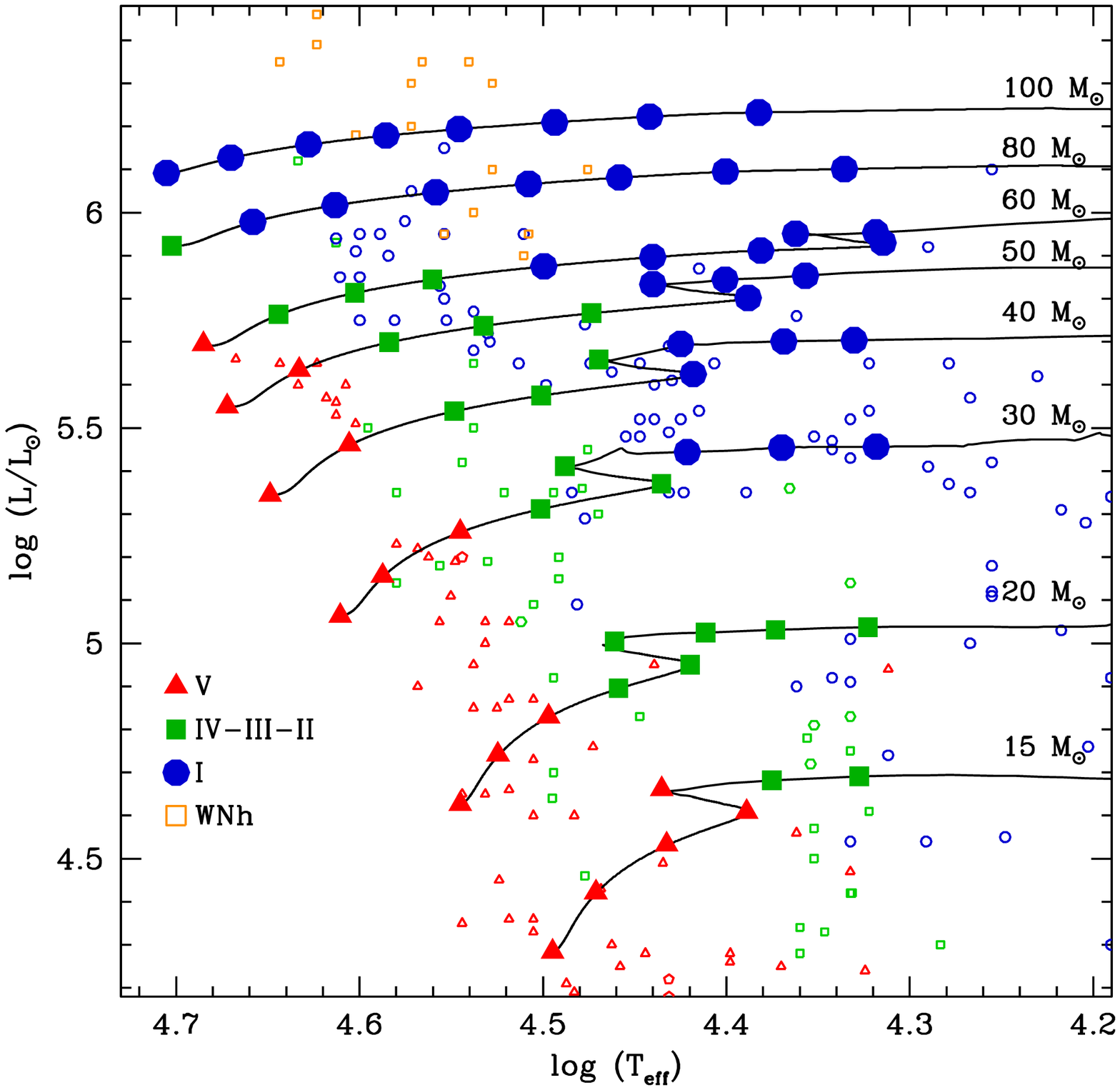}
\includegraphics[width=0.49\textwidth]{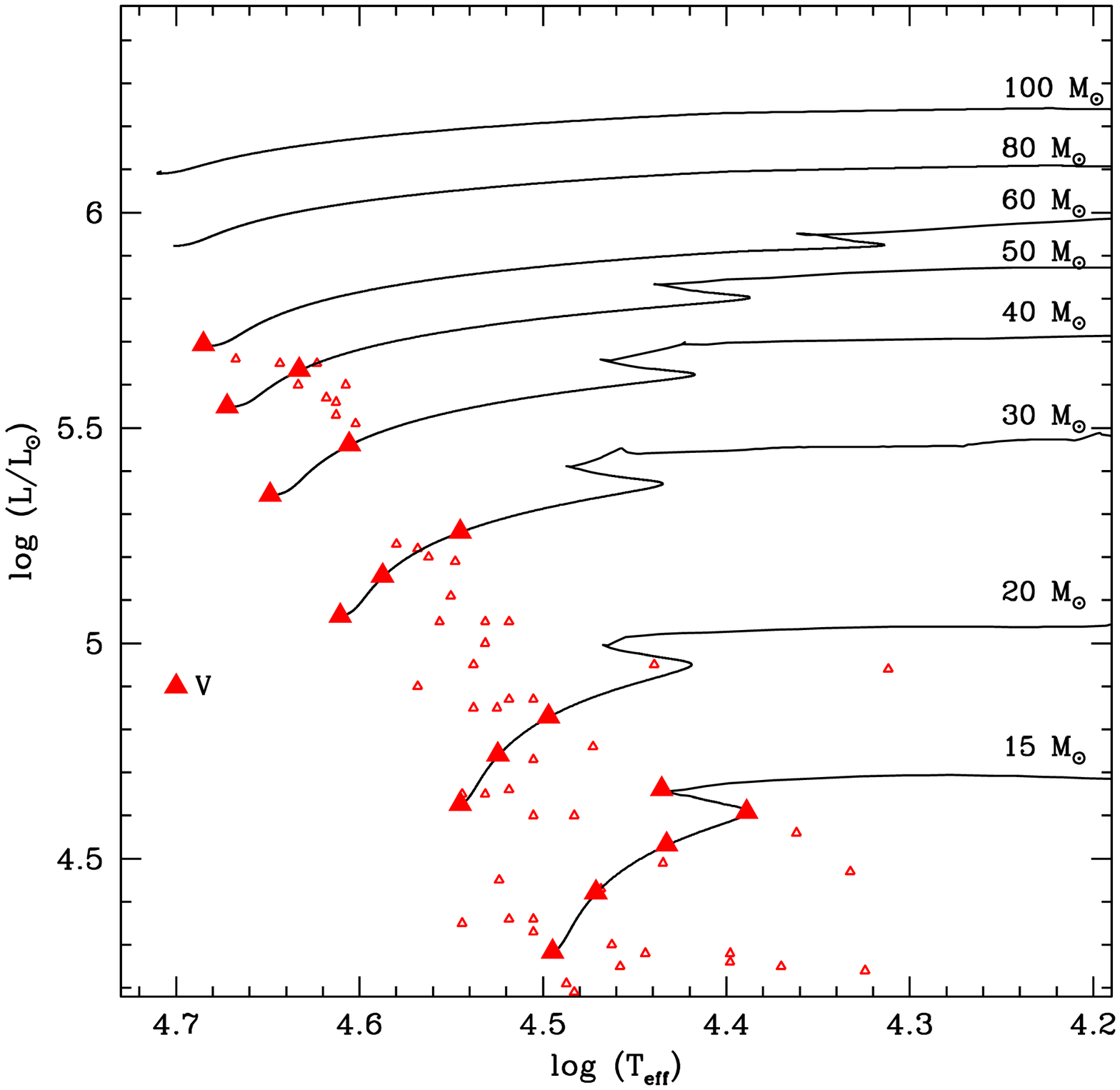}

\includegraphics[width=0.49\textwidth]{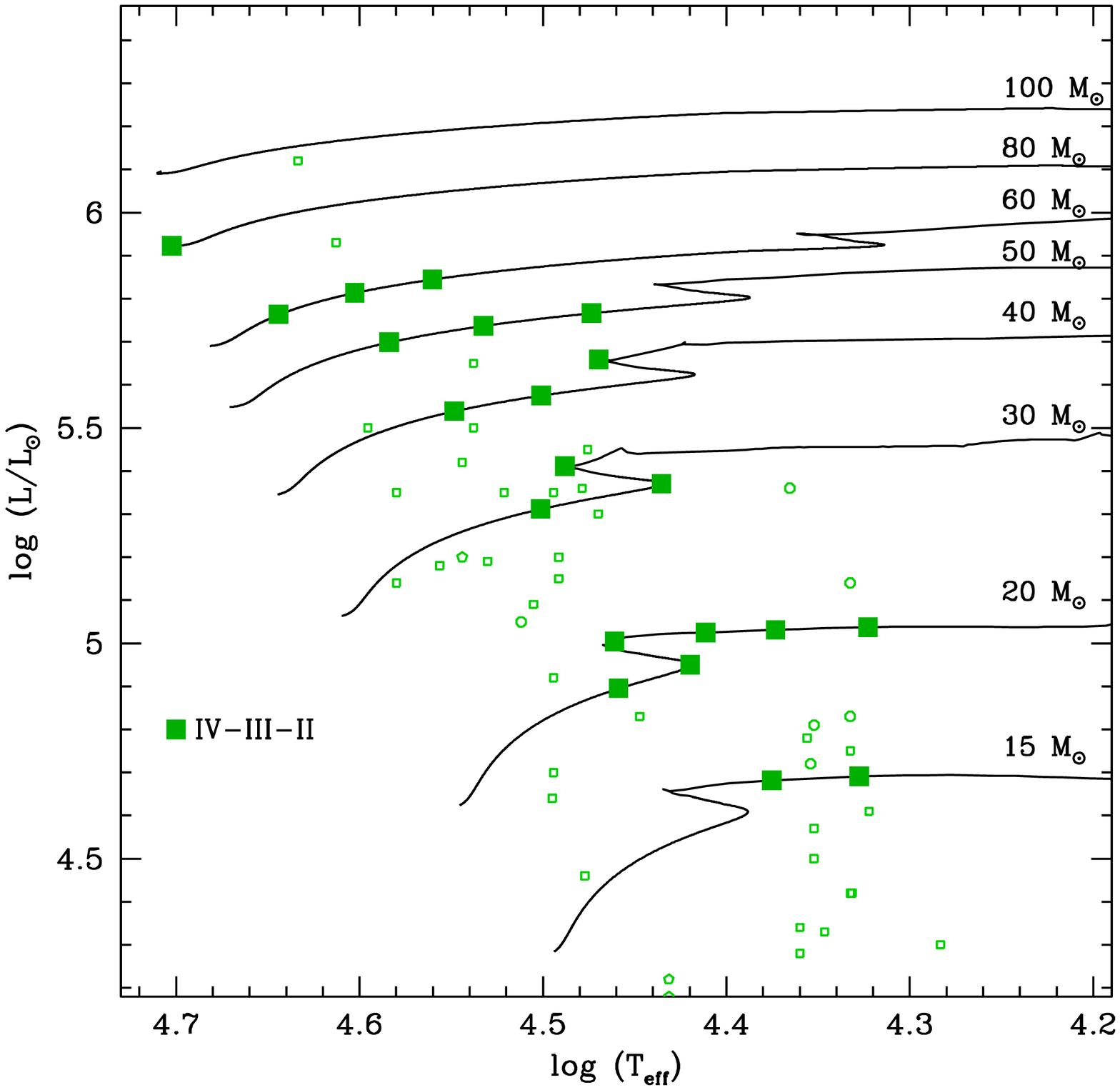}
\includegraphics[width=0.49\textwidth]{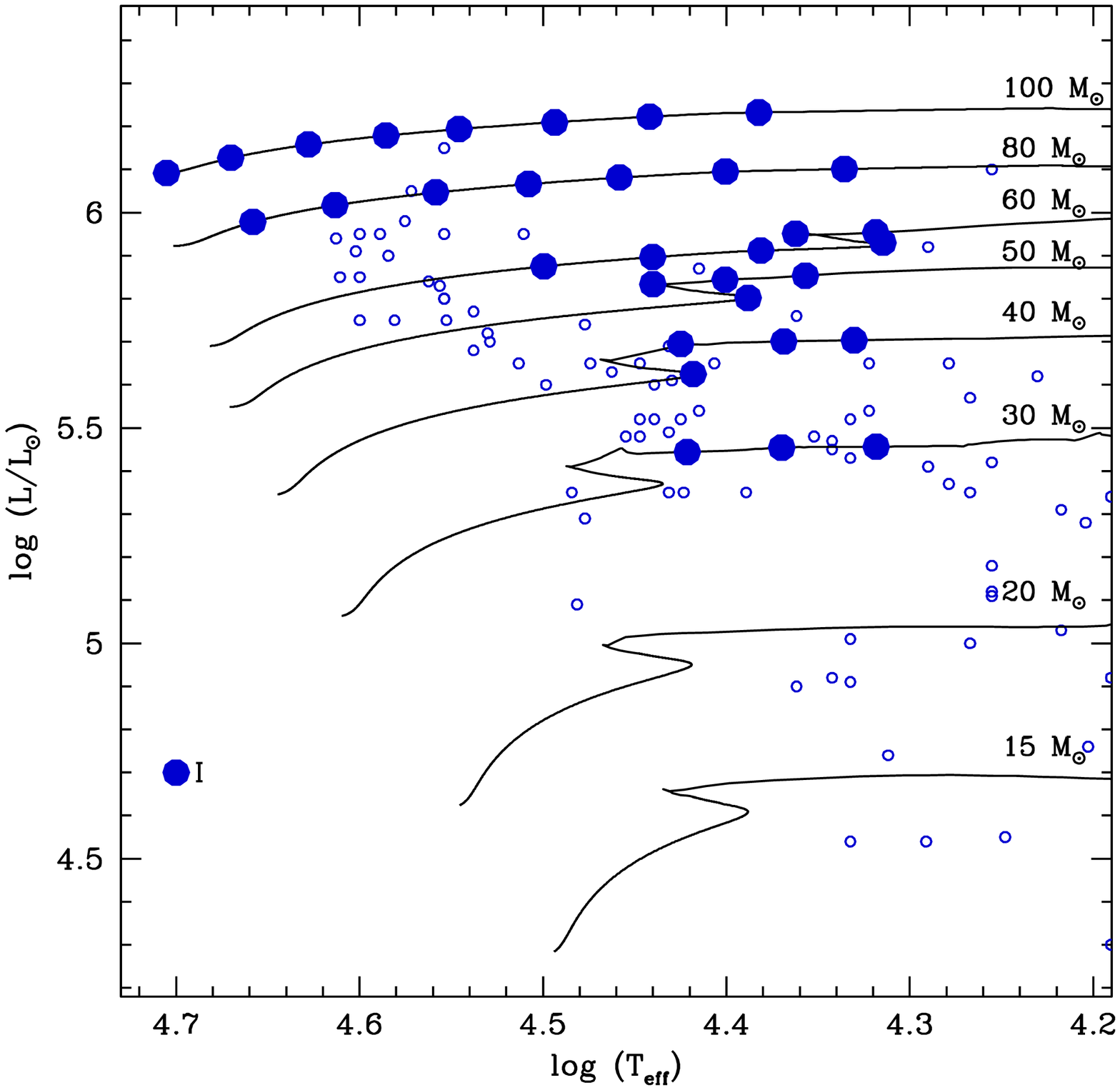}
\caption{Same as Fig.\ \ref{hr_svol} with results from spectroscopic analysis of Galactic stars shown by small open symbols (see list in Sect.\ \ref{s_obs_hrd}). The upper left panel shows the HRD for all luminosity classes, while the upper right (lower left; lower right) shows only dwarfs (subgiants/giants/bright giants; supergiants).}
\label{hr_svol_compobs_init}
\end{figure*}

In Fig.\ \ref{hr_svol_compobs_init} we show our sequences of luminosity classes together with results from spectroscopic analyses using atmosphere models. The latter are taken from  \citet{mcerlean99},  \citet{rep04}, \citet{ww05}, \citet{paul06bsg}, \citet{searle08}, \citet{mark08}, \citet{arches}, \citet{hunter09}, \citet{paul10}, \citet{lef10}, \citet{przy10}, \citet{ngc2244}, \citet{jc12}, \citet{nievaprzy14}, \citet{varnarval}, and \citet{mahy15}. In the following, we refer to these results as the ``observed stars''. Looking at dwarf stars (upper right panel), we see that the distribution of our predicted luminosity classes agrees well with that of observed stars. At M = 15 \msun, the length of the main sequence matches the distribution of predicted and observed dwarfs well. As mass increases, however, the distribution of dwarfs is increasingly more restricted toward a short portion of the main sequence, close to the ZAMS. Hence, above 15 \msun\ , the distribution of dwarf stars in the HR diagram is not an indicator of the main-sequence width. Consequently, such a distribution cannot be used to constrain the size of the convective core (and the related amount of overshooting). 
Another interesting feature in Fig.\ \ref{hr_svol_compobs_init} (upper right panel) is the absence of dwarfs above \lL\ $\sim$ 5.7. The ZAMS models for the 80 and 100 \msun\ tracks already appear as giants or supergiants (see lower panels). This characteristic is further discussed in Sect.\ \ref{s_Vmax}.

The position of giant stars (lower left panel of Fig.\ \ref{hr_svol_compobs_init}) from our synthetic spectra reproduces that of observed giants
reasonably well, especially given that observed giants and dwarfs
or supergiants somewhat overlap in the HR diagram. Regarding supergiants (lower right panel of Fig.\ \ref{hr_svol_compobs_init}), the observed stars behave qualitatively in the same way as our predictions: the higher the mass, the closer to the ZAMS the appearance of supergiants. Below 40~\msun, supergiants are seen only for effective temperatures cooler than that of the TAMS. Above that limit, they are seen already on the main sequence and even on the ZAMS for M=100~\msun. Quantitatively, the observed and predicted distributions are somewhat different in the sense that the former starts at hotter temperatures than the latter, in the mass range 40-80 \msun. \\

In conclusion, our synthetic spectra reproduce the distribution of O stars in the HR diagram relatively well, especially in the hotter part of the main sequence. The quantitative difference in the distribution of supergiants may be mainly attributed to mass loss. Indeed, luminosity classification criteria depend on \ion{He}{ii}~4686, which is sensitive to the wind density (see Sects.\ \ref{s_classif} and \ref{s_compGroh}).

%--------------------------
\subsection{Maximum mass for O-type dwarf stars}
\label{s_Vmax}

We have seen in Fig.\ \ref{hr_ST} and Table \ref{tab_mod} that above 60 \msun\ the luminosity class V was not attributed: the first model of the 80 \msun\ track is a giant. On the 60 \msun\ track, the ZAMS model is classified as O3V((f*)), and it is the only one along that track to have a luminosity class V. 
When mass decreases, more and more dwarf classifications appear along the main sequence. This is indicative of a maximum mass above which luminosity class V is not seen any more.

\begin{figure}[]
\centering
\includegraphics[width=0.47\textwidth]{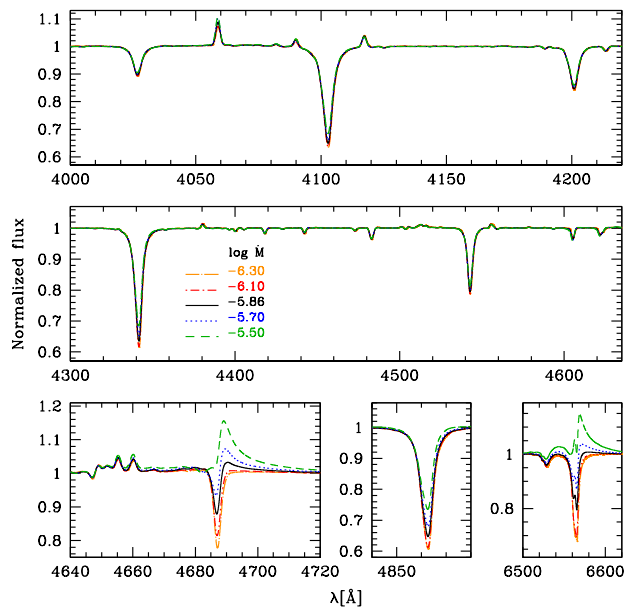}
\caption{Comparison between the synthetic spectrum of the M = 80 \msun\ ZAMS model (black solid line) and the same model with the mass-loss rate decreased or increased by factors 1.4 to 2.8. The mass-loss rate has only been changed in the atmosphere model for these computations.}
\label{m80_wind}
\end{figure}

Inspection of the GOSSS catalog \citep{sota11,sota14} reveals that the earliest O dwarf in the Galaxy has a spectral type O3V((f*)) (HD~64568). The only O2V star known in the Milky Way was recently
reported by \citet{rl16}. The GOSSS catalog does not contain any O2III star either, and only three O2-2.5I stars. O2 stars are thus extremely rare in the Galaxy, and mostly appear as supergiants. In the lower metallicity environment of the Large Magellanic Cloud, \citet{walborn14} identified seven O2 stars in 30~Doradus, five of them being dwarfs (a sixth one having an uncertain luminosity class V-III). Lower metallicity stars being hotter for a given spectral type \citep{massey04,massey05,massey09,mokiem06}, one naturally expects a higher occurence of O2 stars in the Magellanic Clouds.

Differences in the luminosity class distribution can be explained by wind effects. \citet{mokiem07} performed a quantitative analysis of the stellar and wind properties of two O2V((f*)) stars in the LMC, BI~237 and BI~253,  and obtained \lL\ = 5.83 and 5.93, respectively\footnote{\citet{rg12b} found \lL\ = 5.83 and 5.97.}.  In Fig.\ \ref{hr_svol}, these two objects, with these luminosities and effective temperatures of $~53000$ K, would lie around the 80 \msun\ track, consistent with the masses of 75 and 84 \msun\ obtained by \citet{mokiem07}. In our calculations, a star at that position appears as an O2III(f*). The spectral type is thus the same (O2), but the luminosity class is different. Since the latter is driven by the morphology of \ion{He}{ii}~4686, and since \ion{He}{ii}~4686 is sensitive to wind density, we conclude that the stronger mass-loss rate at higher metallicity \citep{mokiem07b} explains that O2 dwarfs are rarely seen in the Galaxy. A test of this hypothesis is shown in Fig.\ \ref{m80_wind}. We computed several models with reduced or increased mass-loss rates compared to the ZAMS 80~\msun\ model. In particular, the model shown by the red dot-dashed line corresponds to a scaling $\dot{M} \propto Z^{0.8}$ assuming $Z = 0.5 Z_{\odot}$. Based on the morphology of \ion{He}{ii}~4686, we would classify it as  O2V((f*)). Figure\ \ref{m80_wind} also reveals that models with high mass-loss rates (higher metallicity) have \ion{He}{ii}~4686 in emission, consistent with a luminosity class I. This confirms that early-type O dwarfs are more easily seen at low metallicity, and supports the fact that very early O dwarfs are almost unobserved in the Galaxy. 

Similar conclusions were reached by \citet{paul10}. In their Fig.~11 they show synthetic spectra of their ZAMS 85 \msun\ model using various mass-loss rates. They discuss these spectra in the context of the nature of the most massive stars. They find that under classical approximations regarding mass loss and clumping (Vink et al. \mdot\ recipes and a clumping factor of 0.1), no dwarf is observed on the ZAMS above 85 \msun\ , that is, at solar metallicity.

%--------------------------
\subsection{Onset of WNh stars}
\label{s_wnh}

In Fig.\ \ref{hr_svol_compobs_init} we have included results from the spectroscopic analysis of WNh stars from \citet{arches} and \citet{paul10}. Such stars are believed to be O-type stars that are still on or close to the main sequence (i.e., showing products of CNO burning at their surface) but having strong winds,
which causes them to appear as Wolf-Rayet stars. A striking conclusion from Fig.\ \ref{hr_svol_compobs_init} is that our models do not reproduce these stars: we predict only supergiants, but no WNh stars.

\begin{figure}[]
\centering
\includegraphics[width=0.49\textwidth]{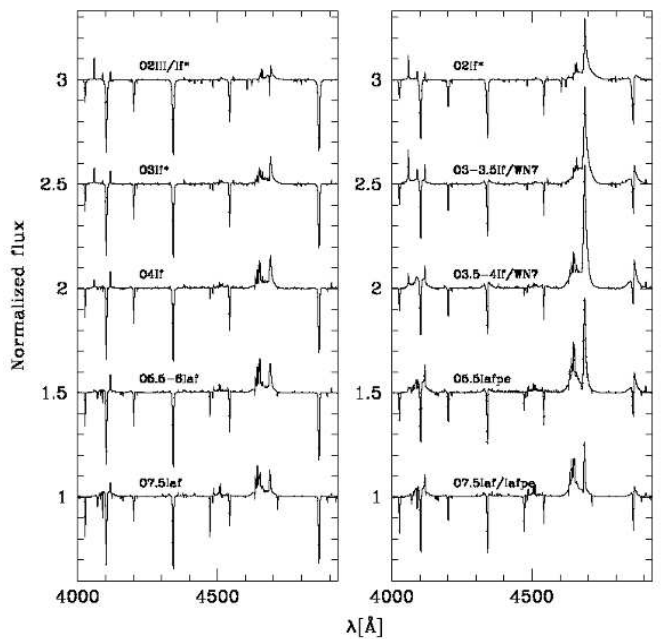}
\caption{Effect of mass loss on the spectral appearance of the 100 \msun\ track. \textit{Left panel}: subset of the initial sequence. \textit{Right panel}: models from the track with a mass loss increased by a factor of 3 compared to the initial sequence.}
\label{m100_wind}
\end{figure}

\begin{figure}[]
\centering
\includegraphics[width=0.49\textwidth]{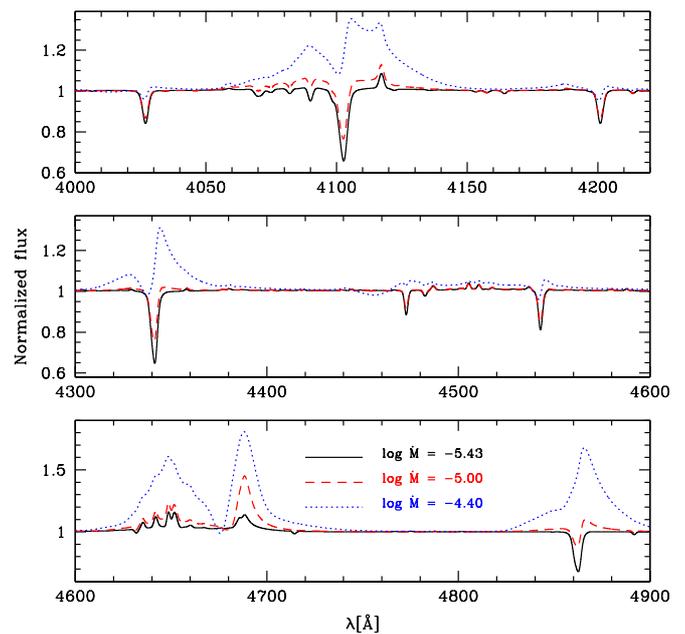}
\caption{Effect of a change in mass loss on the spectral appearance of the fourth model of the 100 \msun\ track. Mass loss was only
modified in the atmosphere model, not in the evolutionary model. }
\label{m100_4_mdot}
\end{figure}

A first possibility to remedy this situation is to assume that the mass-loss rates of WNh stars are higher than assumed in our calculations. In Fig.\ \ref{m100_wind} the models in the right panel correspond to mass-loss rates increased by a factor of 3.0, consistent with the recipe of \citet{vink01} -- $\log \dot{M} \sim\ -5.0$. This is not enough to produce the spectral features of WNh stars. Such stronger winds do lead to the appearance of transition objects (O3-4If/WN7 stars) characterized by \ion{He}{ii}~4686 emission and H$\beta$ with a P-Cygni profile, as defined by \citet{cw11}. Pure H$\beta$ emission typical of WN7h stars is not observed,
however.
The values determined by \citet{arches} and \citet{paul10} for the WNh stars in Fig.\ \ref{hr_svol_compobs_init} are between -5.0 and -4.30. In Fig.\ \ref{m100_4_mdot} we show three models for the fourth point along the 100 \msun\ track. The black and red lines correspond to the models with the initial mass-loss rate and \mdot\ increased by a factor of 3.0, respectively. The third model is a new one with $\log \dot{M} = -4.4$. Its other parameters are the same as that of the initial model. It shows features that are typical of WN6-7ha stars \citep[see Fig.\ 1 of][]{cw11}. Hence, to reproduce the spectral features of WNh stars, mass-loss rates in excess of the prescriptions of \citet{vink01} by a factor of $\sim$ 4.0 (combined with a clumping factor of 0.1) are necessary. However, this is not without problems. If we increase \mdot\ by such a factor in the evolutionary models, then the track we obtain does not reach the position of WNh stars. The mass removal due to winds is so strong that the track almost immediately turns down to lower luminosities. This implies that it is not possible to obtain a consistent evolutionary and atmosphere model with M=100~\msun\ that accounts for WNh stars, at least under the assumption that WNh stars are objects on or just past the main sequence.

A way out of this discrepancy is to assume that the observed luminosity of the WNh stars used for our comparisons may be underestimated. \citet{paul10} found that the stars in NGC~3603 had \lL\ between 6.2 and 6.4, higher than our 100 \msun\ track. They also discusses some of the stars of \citet{arches}, arguing that with different assumptions on the distance and extinction, these objects would be more luminous. 

Another possibility is to assume that the mass-loss rates are increased only after the star has evolved off the ZAMS, at the position where WNh stars are observed. \citet{best14} report that $\log \dot{M}$ scales linearly with $\log \Gamma_{e}$ ,
where $\Gamma_{e}$ is the electron scattering Eddington factor, but with a steeper slope at larger $\Gamma_{e}$, that is, closer to the Eddington limit. We may then speculate that massive stars start with ``normal'' mass-loss rates and then experience an increased mass loss when evolving closer to the Eddington limit. An envelope inflation may also appear close to the Eddington limit, causing problems in our method of combining atmosphere and evolutionary models (see Sect.\ \ref{slim}).

%--------------------------
\subsection{Limitations of the present study}
\label{slim}

The attempt made in this study to provide a spectroscopic view of the evolution of massive stars near the main sequence suffers from several limitations and approximations. The first is that the coupling between evolutionary and atmosphere models remains crude. We ensured that the temperature structures of both models were consistent in the overlapping region (outer part of the evolutionary model - inner atmosphere). This remains true as long as the stars do not develop to large envelope expansion. As luminosity and radius increase, our approach is progresively less consistent. This may explain part of the difficulties we encounter in reproducing WNh stars. In any case, the atmosphere models used as a boundary condition for the evolutionary calculations are approximate. The effect of more accurate atmospheres remains to be tested. 

For luminosities higher than $10^6 L_{\odot}$ and effective temperatures lower than $\sim$ 20000 K, no supergiants are observed (Fig.\ \ref{hr_svol_compobs_init}). This is the $\Omega-\Gamma$ limit \citep{mm00} above which stars are expected to be unbound due to radiative pressure and rotation. Our models extend beyond this limit, populating the upper right part of the HR diagram. This is a clear limitation of our evolutionary models and is  mainly due to the mass-loss prescription used in our computation: we do not take into account the high mass-loss rates encountered in the luminous blue variable phase. Hence, the redward extent of our most massive models should not be regarded as fully realistic.

A different value of the core-overshooting parameter would modify the size of the convective core and the luminosity in the post-main-sequence evolution. From Fig.~1 of \citet{mp13}, we conclude that the inclusion of overshooting leads to a typical increase in \lL\ by 0.05 dex. According to \citet{vink01}, this translates into a change in mass-loss rate by 0.1 dex. Hence, the wind-sensitive lines may be slightly affected. However, we do not anticipate this effect to drastically change our conclusions. 

A major limitation of our work is the assumption that the theoretical spectra we produce correctly reproduce the spectral features observed in massive stars, so that they can be fully trusted when assigning spectral types and luminosity classes. The \ion{He}{i}~4471 and \ion{He}{ii}~4542 lines used as diagnostics of O-type stars spectral types are well reproduced by models in all recent studies and thus should not cause any bias in our results. Additional criteria based on \ion{He}{i}~4388 and \ion{He}{i}~4144 were defined by \citet{sota11} for O8.5-B0 stars. These lines are known to be more sensitive than \ion{He}{i}~4471 to details of the modeling (blanketing, microturbulence), as demonstrated by \citet{paco06}. Although recent studies show that they can be well fitted \citep[e.g.,][]{mark08,jc12,mimesO}, this limitation must be kept in mind. For early-B stars, the main spectral-type criterion is the relative strength of \ion{Si}{iv}~4089 and \ion{Si}{iii}~4552. The former line is not always quantitatively reproduced in spectroscopic analyses \citep{oc16}, although this is not systematic, as stressed in Sect.\ \ref{s_res} and Fig.\ \ref{m20_I_III} \citep[see also best fits obtained by][]{jc12,mimesO}. Hence, an uncertainty of about one subtype may affect the classification of early-B stars. The main luminosity class diagnostic for O stars is the strength of \ion{He}{ii}~4686. Its strength and morphology depends on the wind parameters, as discussed previously. Provided these parameters are constrained, it is usually well reproduced by models, so that there is no major problem with the modeling of this line. For B stars, \ion{Si}{iv}~4089 and \ion{He}{i}~4388 are involved, among other criteria, in the assignment of luminosity classes, implying an uncertainty larger than for O stars. This is seen in Table \ref{tab_mod} where on average it is more difficult to give a unique luminosity class to B-type than to O-type stars.   

The classification of the earliest O-type stars relies on nitrogen lines \citep{walborn02}. \citet{rg11,rg12} studied the formation of \ion{N}{iii}~4640 and \ion{N}{iv}~4058, which are used to distinguish O2-O3-O3.5 stars. They showed that the intensity of these lines depended on stellar winds and metallicity. Comparison between predictions of the FASTWIND code \citep{puls05} and CMFGEN \citep{hm98} revealed good agreement for \ion{N}{iii}~4640, except in a narrow temperature range (30000-35000 K). At these temperatures, the \ion{He}{i}/\ion{He}{ii} ratio is used for spectral classification. Consequently, the discrepancies in the theoretical predictions for \ion{N}{iii}~4640 do not affect our spectral types / luminosity classes. For \ion{N}{iv}~4058, \citet{rg12} concluded that larger differences between FASTWIND and CMFGEN existed, highlighting the uncertainties in the prediction of its intensity. In parallel, \citet{jc12} analyzed a sample of early-type Galactic O supergiants and often had difficulties to fit \ion{N}{iv}~4058, even when most other lines where correctly reproduced. Hence, we stress that the intensity of \ion{N}{iv}~4048 may not be completely under control in our models. Spectral types in the range O2-O3.5 are thus more uncertain than in the range O4-B1.

In the present study, we have focused on single-star evolution without rotation. Taking into account rotation can lead to two main changes in the prediction of spectroscopic appearance. First, evolutionary tracks for rotating stars are modified: they are less luminous and cooler on the ZAMS, but become rapidly more luminous on the main sequence and in the more advanced phases \citep{mema00}. Second, rotation affects mass loss \citep{mame00} so that the wind density is changed compared to non-rotating models. Hence, for a given initial mass, we can expect modifications in the spectral appearance of evolutionary tracks, both in terms of spectral type and luminosity class \citep{markova11}. A detailed study of the such effects will be presented in a future publication.

%%%%%%%%%%%%%%%%%%%%%%%%%%%%%%%%%%%%%%%%%%%%%%%%%%%%%%%%%%%%%%%%%%%%%%%%%%%%%%%%%%%%%%%%%%%%%%%%%%%%%%%%%%%%%%%%%%%%%%%%%%%%%%%
%%%%%%%%%%%%%%%%%%%%%%%%%%%%%%%%%%%%%%%%%%%%%%%%%%%%%%%%%%%%%%%%%%%%%%%%%%%%%%%%%%%%%%%%%%%%%%%%%%%%%%%%%%%%%%%%%%%%%%%%%%%%%%%
\section{Conclusions}
\label{s_conc}

We have presented a spectroscopic view of massive stars evolution at solar metallicity and without rotation. To do this, we have computed evolutionary models with the code STAREVOL for initial masses equal to 15, 20, 30, 40, 50, 60, 80, and 100 \msun. For selected points on these tracks, we computed atmosphere models and synthetic spectra with the code CMFGEN. The resulting optical spectra were classified as if they were observed spectra to provide a spectral type and luminosity class. We obtained theoretical evolutionary sequences. The main results are listed below.

\begin{itemize}

\item[$\bullet$] The earliest spectral types (O2 to O3.5) are only obtained for stars more massive than about 50 \msun. For later spectral types, lower masses are possible. 

\item[$\bullet$] A luminosity class V does not correspond to the entire main sequence above 20 \msun. Dwarf stars are observed all the way from the ZAMS to the TAMS only for the 15 \msun\ track. As mass increases, an increasingly larger portion of the main sequence is spent in the luminosity class III, and dwarfs are not seen on the main sequence above 80 \msun. Supergiants (luminosity class I) appear before the end of core-hydrogen burning above 50 \msun. Consequently, the distribution of luminosity class V stars does not trace the main sequence and cannot be used to constrain the size of the convective core. 

\item[$\bullet$] The distribution of luminosity classes in the HR diagram reproduces the position of dwarfs, giants, and supergiants well as determined by quantitative spectroscopy. A slight discrepancy exists for supergiants and is attributed to wind density. This general agreement indicates that mass-loss rates reduced by a factor of 3.0 compared to the theoretical predictions of \citet{vink01} are a relatively good description of O star winds, provided a clumping volume-filling factor of 0.1 is adopted. 

\item[$\bullet$] We predict an upper mass limit of 60 \msun\ for dwarfs at solar metallicity. This is consistent with the rarity of O2V stars in the Galaxy. This mass limit increases at lower metallicity. 

\item[$\bullet$] Luminous WNh stars are not predicted by our most luminous models. Higher mass-loss rates would be needed to produce strong emission lines, but such rates would yield evolutionary tracks that do not reach the observed luminosities of WNh stars. Their luminosities may be underestimated, so that they may be stars more massive than 100 \msun. Alternatively, they might experience an increase in mass loss as they evolve closer to the Eddington limit.

\end{itemize}

%%#####################################################################
\section*{Acknowledgments}

We thank John Hillier for making his code CMFGEN available to the community. We thank Jes\'us Ma\'iz-Apell\'aniz for developing and maintaining the GOSSS catalog of Galactic O stars, and Evelyne Alecian for sharing the spectrum of HD~122451. We acknowledge financial support from the Agence Nationale de la Recherche (grant ANR-11-JS56-0007). We thank the referee, Alex de Koter, for a detailed and constructive report.

%%#####################################################################
\bibliographystyle{aa}
\bibliography{ev_MSwidth}
%%#####################################################################
%%#####################################################################

\end{document}